\documentclass{sig-alternate-10pt}

\usepackage[sort,square,comma,numbers,compress]{natbib}
\usepackage{pgf}
\usepackage{tikz}
\usepackage{amssymb,amsmath} 
\usepackage{graphicx}
\usepackage{tabularx}
\usepackage{courier}
\usepackage{bm}
\usepackage{textcomp}

\usepackage{fancyhdr}
\usepackage{lipsum}

\usepackage[caption=false]{subfig}
\usepackage[font=small,labelfont=bf]{caption}

\usepackage{algorithm, algorithmic}
\usepackage{url}

\date{}

\author{Greg Kuperman, Jun Sun, Bow-Nan Cheng, Patricia Deutsch, and Aradhana Narula-Tam   \\ 
MIT Lincoln Laboratory\\Lexington, MA, USA 02420 \\ \{gkuperman, jun.sun, bcheng, patricia.deutsch, arad\}@ll.mit.edu }

\title{Group Centric Networking: A New Approach for Wireless Multi-Hop Networking to Enable the Internet of Things}

\begin{document}

\maketitle 

\begin{abstract}
In this paper, we introduce a new networking architecture called Group Centric Networking (GCN), which is designed to support the large number of devices expected with the emergence of the Internet of Things. 
GCN is designed to enable these devices to operate collaboratively in a highly efficient and resilient fashion, while not sacrificing their ability to communicate with one another. 
We do a full protocol implementation of GCN in NS3, and 
demonstrate that GCN utilizes up to an order of magnitude fewer network resources than traditional wireless networking schemes, while providing high connectivity and reliability.
\let\thefootnote\relax\footnotetext{This work is sponsored by the Defense Advanced Research Program Agency via Air Force contract \#FA8721-05-C-0002. 
The views, opinions, and/or findings contained in this article
are those of the authors and should not be interpreted as representing the
official views or policies of the Department of Defense or the U.S.
Government.
Approved for public release; distribution unlimited.
}
\end{abstract}

\section{Introduction}
\label{sec:introduction}

Despite decades of effort, multi-hop wireless networks have not succeeded in fulfilling their once-promised potential of providing ubiquitous connectivity with minimal fixed infrastructure. 
Today, almost all of our wireless devices are still tethered to wired infrastructure such as cell towers or 802.11 access points. 
But with the forecasted explosion in terms of users and data rates \cite{50billion,beyondgigabit,index2011global}, having all devices directly connected to fixed infrastructure will no longer be tenable: 
wired access points will be overwhelmed 
and will quickly become bottlenecks in the network.
If the concept of the Internet of Things \cite{atzori2010internet, vasseur2010interconnecting,lee2011internet} is taken to its natural extent, then almost \emph{everything} will be a ``smart-device'', with all of these devices being wirelessly connected and exchanging data.
Due to this impending surge of wireless devices, there has been a renewed focus on multi-hop networking to facilitate communications among these devices. 
The Internet Engineering Task Force (IETF) and preliminary 5G standards organizations have already begun putting forth ideas for designing future wireless systems, and multi-hop networking is a cornerstone for many of these next-generation architectures \cite{IETFroll,sheng2013survey,osseiran2014scenarios,watteyne2011manet}. 

In this paper, we propose a new networking architecture called Group Centric Networking (GCN), 
which is designed specifically to enable future networks of these smart-devices to operate collaboratively in a highly efficient and resilient fashion.
In these emerging networks, the devices that will be deployed will be resource limited and will be expected to operate in a lossy environment~\cite{watteyne2011manet}.
Preliminary requirements dictate that any future networking protocol designed for these networks must be scalable and provide high reliability \cite{brandt2010home,dohler2009routing,martocci2010building,pister2009rfc}.
GCN provides scalable connectivity in lossy environments, while not sacrificing the devices' ability to communicate with one another. 
In particular, we design GCN to (1) efficiently handle the various types of traffic that future networks of smart-devices will carry, and to (2) take advantage of the wireless medium to resiliently and efficiently connect the devices of the network. 

Most networking schemes today are designed to support an \emph{address}-centric network, where one user typically acts as a client, and another as a server (e.g., your personal computer as the client, and a video-streaming service as the server). 
These two users can live anywhere in the network, and a network routing protocol establishes a path between the two. 
In future networks, 
this point-to-point routing scheme may no longer be appropriate. 
As others have suggested \cite{kopetz2011internet,dohler2009routing,atzori2010internet}, 
a potential future network might connect a large number of wireless smart-devices that are designed to work together in a local environment to improve the quality of life for a human end-user, or to improve production in a factory.
Instead of potentially long-distance point-to-point connections, 
 these smart-devices will desire to communicate locally within groups to accomplish tasks collaboratively.
We label any network with the above characteristics as being  \emph{group}-centric, where the predominant traffic pattern is for data to be disseminated between a group of devices operating in some local area. 

In particular, a group-centric network has the following characteristics:
\begin{enumerate}
\vspace{-.05in}
\item Devices will be grouped by an inherent set of ``interests'' that are dependent on the tasks they are performing, and these group members will wish to communicate reliably between one another. Devices are not limited to a single group, and can belong to multiple groups.
\vspace{-.05in}
\item The majority of message exchanges will be within some local area, and long-distance traffic will only be a small fraction of overall communications.
\vspace{-.05in}
\item Any device can be a source or a sink, and traffic patterns between them may be one-to-one, one-to-many, many-to-one, or many-to-many.
\vspace{-.05in}
\item Future wireless environments will have a mix of mobile and stationary devices, where mobility will be typically be limited to some local area.
\vspace{-.05in}
\end{enumerate}

An emerging example of a group-centric network is a home or factory automation network, where various sensors, actuators, and controller systems work together to adjust to changing conditions in real-time \cite{brandt2010home,martocci2010building}. 
The devices in these networks will work together to ensure that environmental conditions are correct and that machinery is working properly to facilitate production.
Another example of a group-centric network that extends beyond groupings of low-power smart-devices is a military network, where both movement and communications are inherently localized. 
For movement, military operations are typically restricted to a certain geographic area, and for communications, a recent study shows that 95\% of traffic in military networks travels at most three hops, with only 5\% of traffic being long-range  \cite{ramanathan2010scalability}. 

Current wireless networking schemes are ill-suited to meet the needs of a group-centric network.
Today's approach for multi-hop wireless networking is to create end-to-end routes that are composed of a series of point-to-point links \cite{perkins1999ad,jacquet2003optimized,rajaraman2002topology}.
These schemes are typically modifications of protocols that were designed for wired networks, and
the newest proposals for wireless networking standards continue this link-based routing approach \cite{clausen2012loadng,winter2012rpl,thubert2012objective,goyal2013reactive,shelby2012neighbor,hui2014multicast}.
We believe that the characteristics of the wireless environment inherently make link-based routing unsuitable for wireless group-centric networks.
Any point-to-point wireless link is inherently unreliable due to interference, multi-path, and noise.
The idea of a link is itself borrowed from wired networks:
in a wireless network, there is no one-to-one connection between two radios; transmissions are sent over-the-air and are typically overheard by multiple devices. 
The addition of mobility further degrades link reliability. 
Significant network resources need to be expended to maintain wireless links and routes.

Group Centric Networking eschews links and routes in favor of a scheme designed specifically for the lossy wireless medium. 
The key characteristics of the Group Centric Networking approach are:
\begin{itemize}
\vspace{-.05in}
\item No link state or neighbor information is utilized or maintained, and minimal control information is exchanged. 
\vspace{-.05in}
\item Data is efficiently disseminated only across the region where group members exist. To support this, we develop a novel Group Discovery algorithm that dynamically discovers the region of interest and efficiently selects the minimal amount of relay nodes required to ``cover'' this region. 
\vspace{-.05in}
\item Reliable communications is achieved in an error-prone and mobile environment by using ``tunable resiliency'', where the number of redundant data relays is configurable and is able to self-adjust in response to real-time channel conditions.
\vspace{-.05in}
\item Devices communicate in a many-to-many traffic pattern.  Efficient one-to-one, one-to-many, and many-to-one are subsets. 
\vspace{-.05in}
\end{itemize}

As we will demonstrate, GCN utilizes up to an order of magnitude fewer network resources than traditional wireless routing schemes, while providing superior connectivity and reliability.
We verify our approach by implementing the full set of protocols for Group Centric Networking in NS3 Direct Code Execution (DCE) \cite{ns3,ns3dce}, which allows for a high-fidelity comparison against other wireless networking protocols, and enables an easier transition of GCN protocols to real systems and to other researchers in the community. 
The results were verified in the real-time emulation environment EMANE/CORE \cite{ivanic2009mobile,core_emane}. 

In this paper, we present a new networking approach to connect users in relatively close proximity that share common interests. 
As for any new scheme, we do not solve all problems that may be associated with fully deploying GCN.
In particular, we do not give a full definition of a ``group'' for GCN. 
We envision that there will be a relatively small number of groups in the network, where each group consists of users that will desire to collaborate.  
We leave this topic open for further study.
The outline of this paper is as follows.
In Section~\ref{sec:highlevel}, we present Group Centric Networking and its major mechanisms. 
In Section~\ref{sec:gcn_implementation}, we discuss our implementation of GCN, 
and present simulation results demonstrating its performance. 
In Section~\ref{sec:conclusion}, we conclude and discuss ongoing work and future directions for GCN. 


\section{Group Centric Networking}
\label{sec:highlevel}

In this section, we present the core mechanisms that form Group Centric Networking (GCN). 
GCN is designed to efficiently and robustly support groups of devices or users desiring to communicate with one another in a local region. 
Many new emerging wireless devices will be resource limited and will be expected to operate in a lossy environment.
Hence, communication  must be (1) resilient against packet errors due to interference and mobility, and be (2) bandwidth and power efficient. 

GCN enables a set of users to efficiently and resiliently communicate with any other set of users in a group via a many-to-many traffic pattern.
One user may wish to send data to the entire group or to only some members of that group. 
Alternatively, some set of users may wish to collect information from another set. 
The ``many" of the many-to-many traffic pattern can either be some of the group members, or all of them. 
One-to-one, one-to-many, and many-to-one are all considered subsets of the many-to-many traffic pattern.

\begin{figure}[t]
   \centering
   \includegraphics[width=0.44\textwidth]{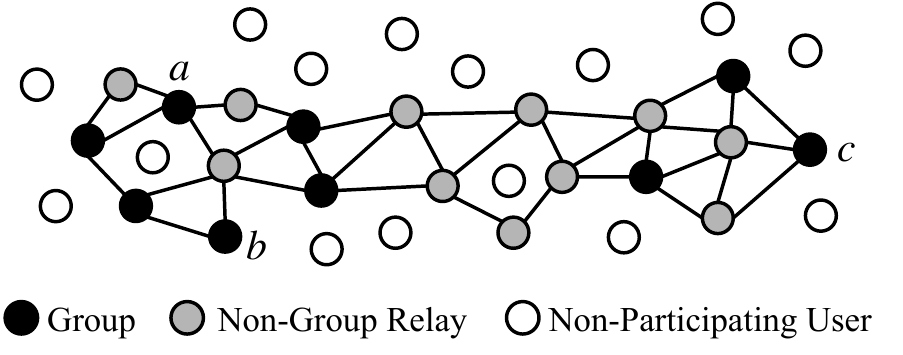} 
   \vspace{-.05in}   
   \caption{An example of a group centric network}
   \label{fig:gcn_example}
   \vspace{-.2in}
\end{figure}

An example layout for a group centric network is shown in Fig. \ref{fig:gcn_example}. 
A set of relay nodes has been activated such that all group members are connected to one another.
There are multiple opportunities to overhear a message in case of packet loss, and the failure of any individual relay or link will not prevent messages from being received by other users. 
User $a$ may wish to communicate to the entire group, or just to users $b$ and $c$. 
Both of these communication types are efficiently enabled by the one-to-many traffic pattern.
Alternatively, $a$ may wish to receive data from users $b$ and $c$ via a many-to-one message. 
While only one group is shown, users can belong to any number of groups. 

To achieve our design goals of scalable, efficient, and resilient group communications, the major mechanisms of Group Centric Networking are as follows:
\begin{enumerate}
\vspace{-.05in} 
\item \emph{Group discovery}: Efficient discovery of the local region where group members reside via a a group discovery algorithm that is able to  connect group members without the use of global control information.
Group Discovery is discussed in Section~\ref{sec:gd}.
\vspace{-.25in} 
\item \emph{Tunable resiliency}: Relay nodes are activated such that the local region is sufficiently ``covered'' in data by having a tunable number of redundant data relays. 
This allows for resiliency towards both packet loss and mobility without the need for the constant exchange of control information.
The number of activated relay nodes self-adjusts in response to real-time channel conditions. 
Tunable resiliency is described in Section~\ref{sec:tr}.
\vspace{-.08in} 
\item \emph{Targeted flooding}: Data can be efficiently and resiliently sent between sets of group members through an approach we call ``targeted flooding''. 
The mechanism for targeted flooding is detailed in Section~\ref{sec:ss}.
\end{enumerate}

\vspace{-.1in}
\subsection{Group Discovery}
\label{sec:gd}

The purpose of group discovery is to find and connect group members in a local region without prior knowledge of where those group members reside, and to do so efficiently without globally flooding control messages.
A naive approach would be to use a control message for discovery that has some time-to-live\footnote{Time-to-live (TTL) is a field used in data packets to limit the distance a packet travels. Each time a packet is retransmitted, the TTL gets decremented by one, and once the TTL reaches zero, the packet is dropped.} 
(TTL) set to the maximum number of hops the group is expected to extend from end-to-end.
This discovery message is then transmitted across the network, with the TTL being decremented at each next user. 
While the message would reach the entire group, it would also travel into areas where group users do not exist. 
In a large network with limited bandwidth, this can be a significant waste of network resources.

\begin{figure*}[t]
\begin{minipage}[t]{0.49\linewidth}	
	   \centering
	   \includegraphics[width=1\textwidth]{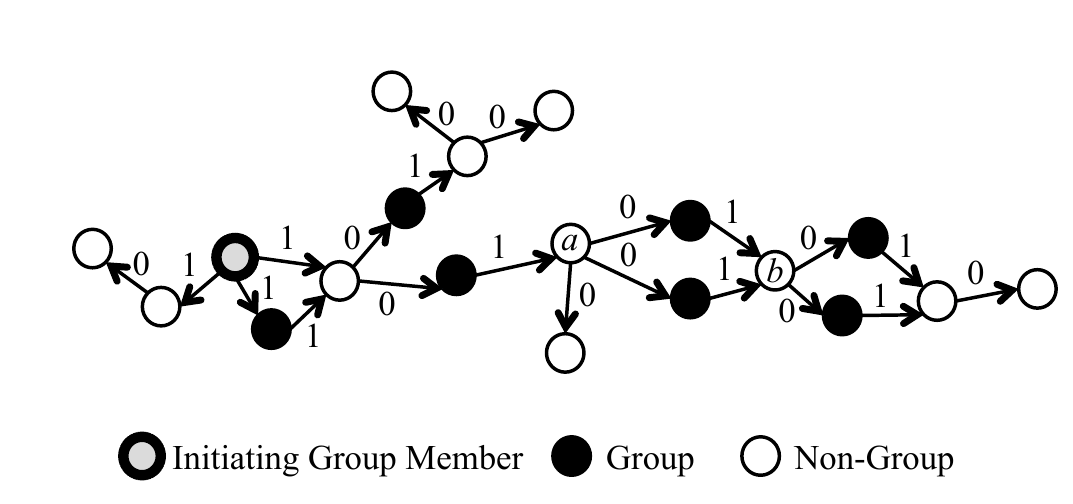} 
	   \vspace{-.25in}   
	   \caption{Discovering the local region using discovery regeneration. Each arrow shows the time-to-live (TTL) of the outgoing discovery message. }
	   \label{fig:gd_1}	
\end{minipage}
\hspace{0.5cm}	
\begin{minipage}[t]{0.45\linewidth}
	   \centering
	   \includegraphics[width=.7\textwidth]{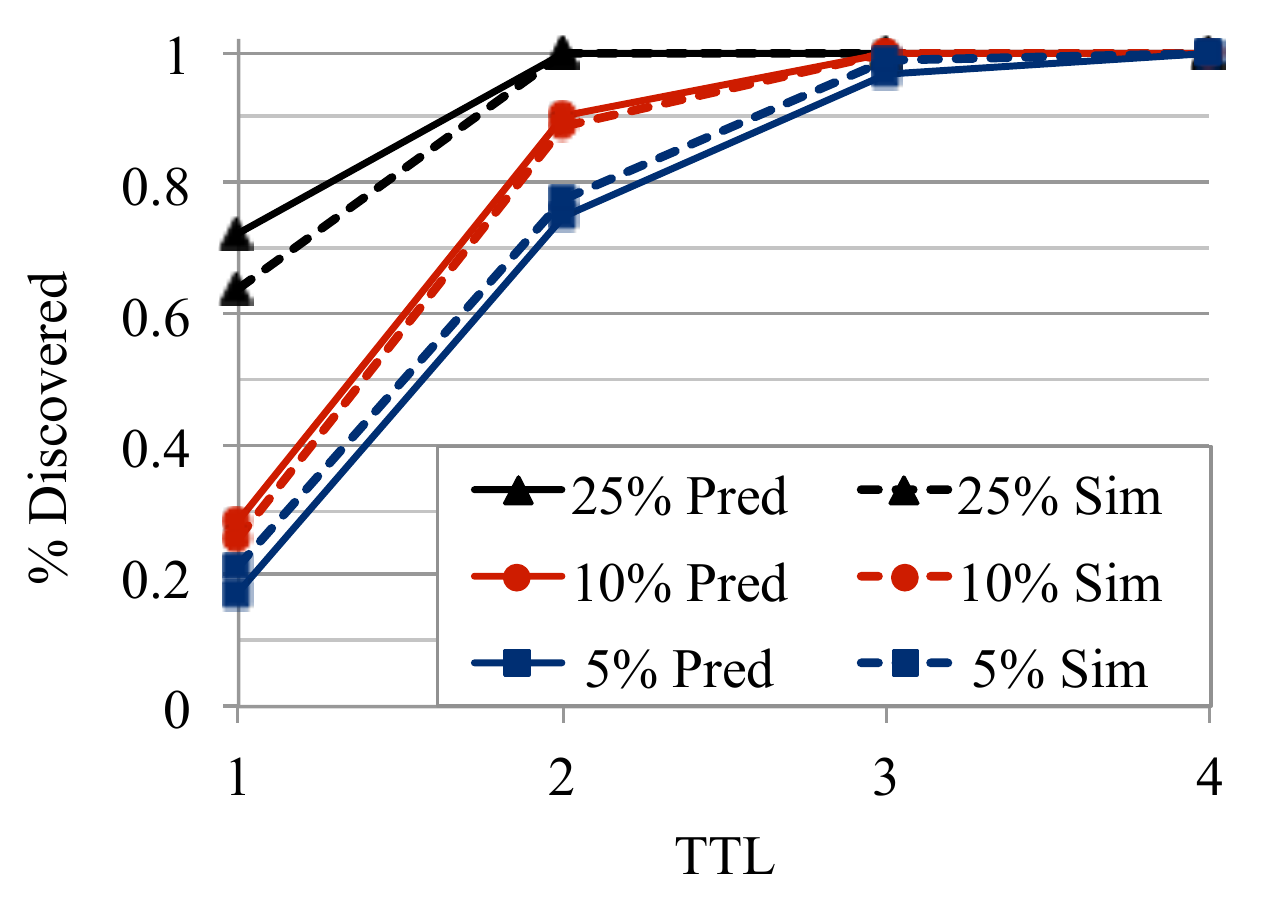} 
	   \vspace{-.1in}   
	   \caption{Predicted and actual percentage of group members found during the discovery process as a function of source TTL.}
	   \label{fig:gd_reach}
\end{minipage}
\vspace{-.2in}
\end{figure*}

For our group discovery algorithm, we introduce a novel approach that we label \emph{discovery regeneration}, where a group discovery message is regenerated with some small ``source'' TTL by each group member. 
By doing so, the reach of the discovery message is limited to some fixed distance around the local region where group members live. 
The basic mechanism for Group Discovery using discovery regeneration is as follows. 
\begin{itemize}
\vspace{-.06in} 
\item A group member initiates group discovery by sending out a discovery message. 
The TTL for the discovery message will be set according to how far around any particular group member the discovery region is to extend. 
We refer to the TTL that the initiating user sets as the \emph{source TTL}.
\vspace{-.06in} 
\item If a group member hears a discovery message, it will regenerate the message with the source TTL.
\vspace{-.06in} 
\item If a non-group member hears a discovery message with a TTL greater than zero, it will decrement the TTL, and rebroadcast the message. 
If a non-group member hears a discovery message with a TTL equal to zero, it does nothing. 
\end{itemize}
\vspace{-.05in} 

Data relays are elected via an acknowledgment (ACK) sent by the group nodes. 
When a group member receives a discovery message, it sends back an ACK 
to the previous group node that relayed the advertisement. 
All nodes in between the group nodes that receive the ACK are elected as data relays. 
If multiple discovery messages are heard, the ACK is sent only to the neighbor that sent the first one. 
Duplicate detection is used by all users in the network to ensure discovery messages are broadcast only once.
Note that ACKs are only sent to the group node that regenerated the source TTL, and not to the source of the initial discovery message. 
This approach is different from traditional multicast whereby {\it join} messages are sent to the root of the tree.

In GCN, when a user becomes activated as a relay, it is now a relay for the entire group, and not for any particular group node.
Relays do not need to maintain any information on who sent it an ACK. 
Similarly, a group member listens for data from any relay, and not only from the relay to which it initially sent an ACK.
After the group discovery process completes, no link-state or neighbor information is maintained by any user in GCN.


An example of group discovery with regeneration is presented in Fig.~\ref{fig:gd_1}. 
Each arrow shows the time-to-live (TTL) of the outgoing discovery message. 
Group members regenerate the TTL at the source value of 2 (which is decremented to 1 at transmission). 
Non-group members do not regenerate the TTL, which limits the reach of the discovery message to the local region where group members reside. 
All group members are discovered without the need to have control information extending  beyond the local area. 
The non-group user $a$ has two group members that hear its discovery message. 
Each of those group members send back an ACK to $a$, and once $a$ receives one ACK, it becomes activated as a group relay node; user $a$ can ignore any additional ACKs that it hears. 
Non-group user $b$ hears a discovery message from two group members, and it will choose one of the two to send an ACK.
By default, $b$ chooses the first user it hears a discovery message from; hence, only one of the two group members that $b$ heard a discovery message from will be activated as a relay. 

\vspace{-.075in}
\subsubsection{Effect of Regenerated TTL on Group Reach}

In order to discover all group members, the regenerated TTL value needs to be sufficiently high such that all users are within the discovery region. 
Setting the TTL too high, however, will result in wasted transmissions in regions where group members do not exist. 
To quantify a recommended TTL value, we perform analysis and simulation to show that low values of TTL are sufficient to discover all group members, even if the group is sparsely populated. 
For the analytic model, we develop a first order approximation that predicts the number of group members that will be discovered as a function of the source TTL. 
Due to space constraints, we only the present the result of our analytic model. 
We consider $N$ users that are uniformly distributed across a two-dimensional region with an area of $A$; the density of users is given by $\lambda=\frac{N}{A}$. 
A user in this region is a group member with probability $P_g$, and  
each user has a transmit distance of $X$. 
Given this set of assumptions, our approximation for the expected percentage of group members that are discovered 
with a source TTL of $T$ is $ 1 -  e^{P_{g} \lambda \pi ((X-\frac{1}{2\lambda}) \cdot T)^{2}}$.


In addition to the analytic model, we performed a simulation using our implementation of GCN in NS3 and compare the results to what is predicted by the analytic model. 
For the simulation, we consider 100 users uniformly distributed in a circular region with a radius of 100 meters, and a transmit distance of 40 meters per user. 
We test three different group densities, where a user is a group member with a probability $P_g$ of either 5\%, 10\%, or 25\%. 
A group member is selected at random to initiate the group discovery process. 
All users are stationary for the duration of the test.
The source TTL is varied between 1 and 4.

Fig. \ref{fig:gd_reach} shows the average simulation and predicted results over 50 random seed runs.
First, we observe that for low values of group membership probability, which leads to group members being far apart, low values of source TTL are sufficient to find all users. 
For a group probability of  5\% (where we expect only five group members on average), a source TTL of 3 allows 98.6\% of the group to be found on average. 
For a group probability of 25\%, a source TTL of 2 finds 99.8\% of group members.  
Next, we observe that the analytic model is a close fit for the results from the NS3 simulation. 
Thus, a user can use the analytic model to select an appropriate source TTL for efficient discovery.


Once the group discovery process is complete, the local region becomes connected via a set of relays that connect the group members to one another. 
A user can now send a message to the entire group via a one-to-all traffic pattern.
The one-to-all pattern forms the backbone of the many-to-many traffic pattern (presented in Section \ref{sec:ss}).
The group discovery process can be periodically rerun to allow new users to join the network that are outside of the local coverage area, or to reconnect users that left the local area because of mobility. 

\vspace{-.075in}
\subsubsection{Total Transmission Comparison}

To understand the savings of transmissions sent over-the-air with GCN (for both data and overhead), we compare it to two different methods for dissemination of data in wireless networks. 
GCN is designed for messaging in a local region; hence, we compare against the flooding scheme of Simplified Multicast Forwarding (SMF)  \cite{macker2012simplified}, which floods a local region with data while employing duplicate packet detection to limit retransmissions.
In SMF, a message is transmitted with some TTL, and that message is then continually rebroadcast by each subsequent user until the TTL expires. 
No control messaging is required in SMF, and there is no mechanism to dynamically set the TTL. 

To offer a fair comparison against GCN, when we operate SMF, we assign a message the minimum TTL for it to reach all of the group members. 
We also compare GCN against the Ad-Hoc On-Demand Distance Vector (AODV) routing protocol \cite{perkins1999ad}, which finds a route from a source to a destination at the time a message is to be sent. 
AODV is used in the ZigBee multi-hop networking standard \cite{baronti2007wireless}, 
and is the basis for new proposals to connect networks of smart-devices \cite{clausen2012loadng,clausen2013lightweight}. 
We note that there exists a multicast version of AODV, called Multicast AODV (MAODV) \cite{royer2000multicast}, but there is no implementation available to compare against. 
In MAODV, each multicast group member requires a unicast route back to the source, 
and to find these unicast routes, MAODV uses AODV control messaging.
Hence, MAODV should utilize the same amount of control traffic as AODV.

In our simulation, group nodes are operating in a local region as consistent with a personal area network.
We consider two concentric circles, one with a radius of 100~meters, and the other with a radius of 200~meters.
Group members reside within the smaller circle, which we denote as the local region.
Users are uniformly distributed across the entire network, with 400 total users (which gives approximately 100 users in the local region).
Each user has a transmit distance of 40~meters.
Of the users in the local region, 10\% are group members (i.e., on average, we expect 10 group members).
All users are stationary for the duration of the test. 
One group member is randomly selected as the source. 
This source will initiate group discovery, and then send a data packet destined to all members of the group at the rate of 1 packet per second for 10 seconds. 
A data packet is 1400 bytes.
In GCN, a group discovery message is 14 bytes and an acknowledgment packet is 20 bytes. 
AODV is run with its default parameters. 

\begin{figure}[t]
	   \centering
	   \includegraphics[width=.35\textwidth]{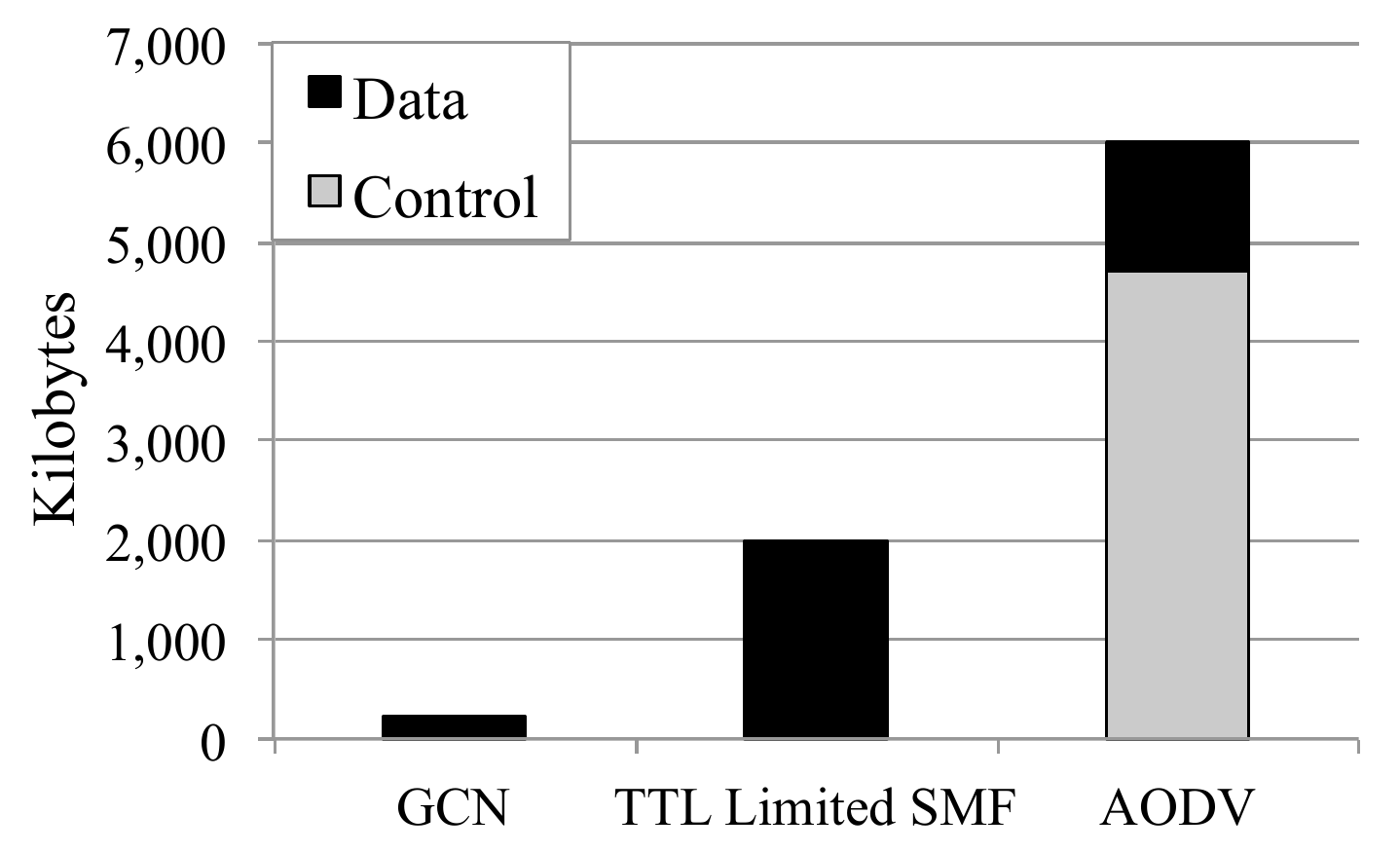} 
	   \vspace{-.05in}   
	   \caption{
		Total bytes sent over-the-air to transmit all data packets from a source to the entire group.
	   }
	   \label{fig:gd_sent}
	  \vspace{-.25in}   
\end{figure}

Fig.~\ref{fig:gd_sent} illustrates the average total traffic sent over-the-air for 50 random seed runs. 
Using the set of relays activated during the group discovery process, GCN is able to efficiently disseminate the set of packets to the entire group. 
GCN transmits a total of 220 KB over-the-air, with only 6.5 KB of that being control information. 
SMF requires 2,001 KB to disseminate these same 10 packets. 
SMF does not discover the local region, and hence floods data into areas where group members do not exist. 
This is particularly problematic when the source node is located at the edge of the group.
AODV transmits a total of 6,000 KB to disseminate the data to the entire group, with 4,700 KB of that total being control traffic.
The reason control traffic is so high is because AODV floods control messages through the entire network, and not just the local area where users are located.
As mentioned earlier, AODV is not a multicast protocol, hence it sends a separate copy of the same packet to each user.
But even if we reduce the data portion by a factor of ten, the control information sent over-the-air is still significantly greater than what either GCN or SMF use in total.

\subsection{Tunable Resiliency}
\label{sec:tr}

As noted above, group discovery activates an efficient set of relays 
such that all group members are connected. 
This immediately enables the one-to-all traffic pattern where a user is able to transmit a packet to the entire group.
But, this minimal set of relays is not particularly robust for group-wide dissemination as a single packet failure can cause all downstream group members to not receive the data.
Additionally, mobility can easily cause the group to become disconnected. 

To make GCN more robust, we extend group discovery by adding a mechanism we call \emph{tunable resiliency},
which 
allows for the targeted activation of additional relays to provide sufficient coverage in order to protect against packet loss and mobility. 
The number of activated relays is able to self-adjust to respond to real-time channel conditions.
To enable tunable resiliency, the group discovery process is extended as follows:
\begin{enumerate}
\vspace{-.05in} 
\item A short delay is added to the discovery acknowledgment (ACK) messages.
\vspace{-.05in} 
\item Each user keeps a count of how many neighbors it sees in order  to determine the number of possible data relays within that user's neighborhood.
\vspace{-.05in} 
\item A set of users will self-select as data relays in a probabilistic fashion to achieve the desired density of relays to enable robust data coverage. 
These probabilistically selected relays are in addition to the set of users that are deterministically selected as data relays through the group discovery process.
\end{enumerate}

The purpose of the short delay before transmitting an ACK is to allow discovery messages to propagate through the immediate vicinity of a particular user. 
Discovery messages are retransmitted as soon as they are received. 

For the neighbor counting process, a user will count the number of discovery messages that it hears from other users in its immediate vicinity. 
When a user receives a discovery message, it then immediately retransmits that message (unless that user is a non-group member and the discovery message has a TTL of zero). 
The neighboring users will receive the discovery message, and immediately rebroadcast it themselves. 
Since the discovery messages are transmitted immediately, after a short amount of time, a user should be able to count the number of discovery messages it has heard.
This allows the user to get an estimate for the number of users that are within its neighborhood. 

By having an estimate on the number of users in a neighborhood, nodes can now self-select as data relays to achieve the desired density for data coverage.
Assume that we wish to have $R$ data relays within range of any particular user.
This value $R$ specifies the density of data relays for the group, and higher values of $R$ provide additional resilience against packet loss and mobility. 
Recall that in the group discovery process, an ACK is addressed to a particular user to activate it as a relay. 
We call this user the \emph{obligate} relay. 
If a user is specified as an obligate in an ACK message, it will always become a relay.
To allow users to self-select as relays, a field is added to the ACK that specifies a probability of accept (ACP). 
If a user receives an ACK and it is not the obligate, it then becomes a relay with probability ACP.
Once a user becomes a relay (either by being the obligate or by self-selecting), it then continues the discovery process by sending a new ACK that follows the same steps as above.
To maintain a uniform distribution of relay nodes across the group region, a user will only attempt to self-select once; i.e., the user will not attempt to self-select a relay with each subsequent ACK it receives.
But if that user is specified as an obligate in any ACK, it \emph{will} become a relay. 

The ACP value is set as follows. 
Assume that a user has counted $N$ neighbors and desires to have a total of $R$ data relays within its range. 
The first node a user hears a discovery message from will be selected as the obligate relay, and the ACP value will be set to $\frac{R-1}{N-1}$. 

This approach for probabilistically selecting data relays allows the network to self-adjust to real-time error conditions.
The number of discovery messages heard by each user reflects the current error rate being experienced in the network. 
For example, assume there is a 50\% packet error rate due to interference or some other loss; if ten neighbors of user $U$ transmit a discovery message, then on average five of those messages should be expected to be heard by $U$. 
If we assume a wireless channel has a similar error rate in both directions (which previous studies show to be typically true \cite{aguayo2004link}), 
then an ACK sent by $U$ should reach a similar number of neighbors that $U$ initially heard a discovery message from (i.e., about five of the ten neighbors should hear an ACK). 
If $U$ desired to have three data relays in its vicinity, it will send an ACK with one obligate and an ACP set to $\frac{2}{4}$. 
On average, this will activate close to the three desired relays.

\begin{figure}[t]
    	\centering
    	\subfloat[Minimum set of relays to connect the group]{
    		\includegraphics[width=0.4\textwidth]{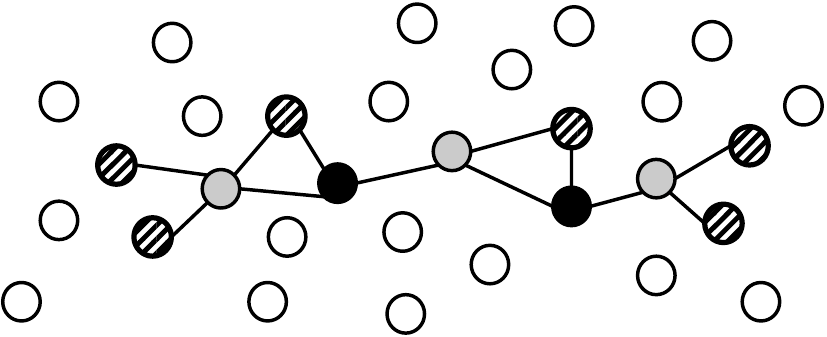} 
    		\label{fig:tr_low}
    	}  	
	\\
    	\subfloat[More relays allows resilient group communications]{
    		\includegraphics[width=0.4\textwidth]{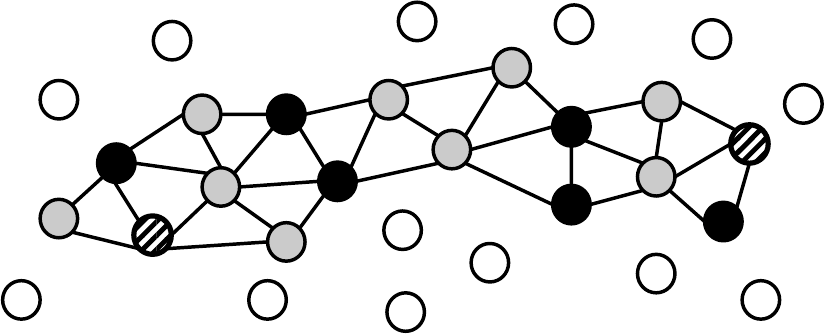} 
    		\label{fig:tr_high}
    	} 	
	\\
	\vspace{-.05in} 
    	\subfloat{
    		\includegraphics[width=.49\textwidth]{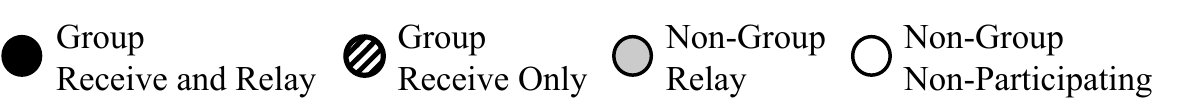} 
    		\label{fig:tr_legend}
    	} 		
	\vspace{-.05in} 
    	\caption{Change in coverage using tunable resiliency. }
    	\label{fig:tr_hl}
	\vspace{-.225in} 
\end{figure}

After an iteration of group discovery with tunable resiliency has been performed, the area where the group resides will have a sufficient density of relay nodes to increase data coverage and become resilient to loss and mobility.

Fig.  \ref{fig:tr_hl} shows an example of tunable resiliency.
In Fig. \ref{fig:tr_low}, the minimum set of relays is activated and the entire group is connected.
When a group member sends a one-to-all transmission, all group members will receive the message.
But, if any packet is lost, or any relay moves out of range, then the group will become disconnected.
In Fig. \ref{fig:tr_high}, additional relays have been activated by setting the ACP  in the ACK message to achieve the target number of relays.
This allows data to cover more of the group area, which increases the resiliency of the group against packet loss and mobility. 

To evaluate the effect that tunable resiliency has on Group Centric Networking, we look at two criteria: (1) the connectivity of a group when users are mobile, and (2) how reliably and efficiently messages can be delivered in the presence of packet loss and mobility.
The desired number of data relays that any user wants to activate is denoted by $R$.

\vspace{-.05in}
\subsubsection{Effect of Mobility on Group Connectivity}

\begin{figure}[t]
   \centering
   \includegraphics[width=0.4\textwidth]{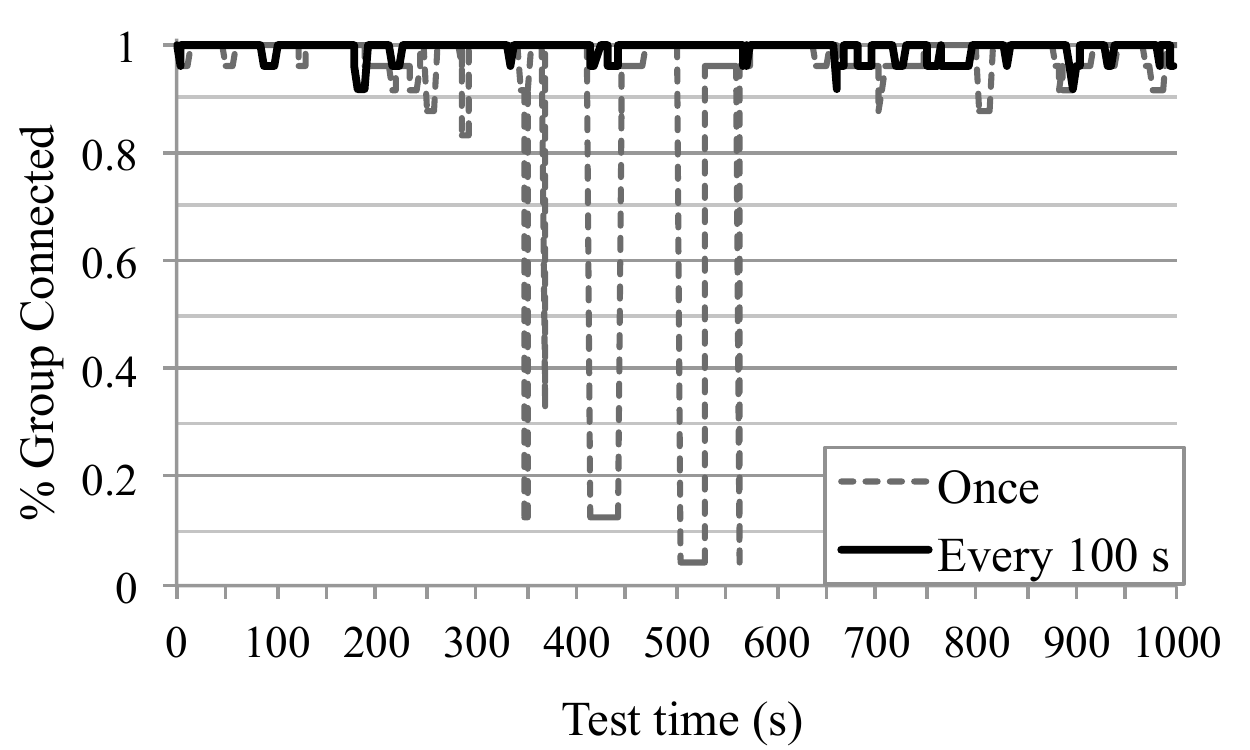} 
   \vspace{-.05in}   
   \caption{Group connectivity with mobile users}
   \label{fig:tr_connected}
   \vspace{-.2in}
\end{figure}

For the first test, we consider 100 users  uniformly distributed  in a circular region with a radius of 100 meters. 
Any user in this region can be a group member with a probability of 25\%. 
The number of desired data relays $R$ is set to 2. 
Users move according the random waypoint model, with a speed of 0 to 5 m/s and a pause time of 0 to 2 seconds.
The test is run for 1000 seconds.
One group member is randomly picked as the source, and it will initiate group discovery. 
The source  then either (1) never initiates another group discovery for the remainder of the test, or (2) initiates group discovery every 100 seconds. 
To measure group connectivity, we sample the network every second and determine if there exists a path from the source to each of the group members.

In Fig. \ref{fig:tr_connected}, we  plot the percentage of group members that are connected to the source as a function of time.
When a group discovery message is sent once every 100 seconds, the group has very high connectivity. 
Overall, the source has a direct connection 99\% of the time to any other group member.
Next, we see that even when only a single group discovery is performed, the group remains highly connected for long periods of time.
With only a single group discovery performed at the beginning, the source has a connection to any other group member for 92\% of the 1000 second test.

\subsubsection{Effect of Resilience Factor in Lossy Environments}

Next, we examine the effect of tunable resiliency on the reliability of message delivery in the presence of packet loss and mobility. 
The following simulation is performed.
100 users are uniformly distributed in a circular region with a radius of 100 meters. 
Any user in this region is a group member with probability of 25\%. 
Each user has a transmission radius of 40 meters.
We test the following three packet error rates (PER): 0\%, 25\%, and 50\%. 
Users are either stationary or move according the random waypoint model with each user selecting a speed from a uniform distribution between 0 and 5 m/s.
One group member is selected at random to initiate group discovery. 
The number of desired data relays $R$ is set to either 1, 3, or 5.
When $R=1$, only an obligate relay will be selected, and tunable resiliency is effectively turned off.
Each group member sends a packet to the group once every second for 100 seconds. 
The number of packets successfully received at each group member is recorded.
Additionally, the total number of bytes sent over-the-air by all users is recorded.
Similar to the test performed in Section \ref{sec:gd}, we compare against SMF, which floods the data across the entire region.
Fifty random tests are run, with the results averaged.

\begin{figure}[t]
   \centering
   \includegraphics[width=.4\textwidth]{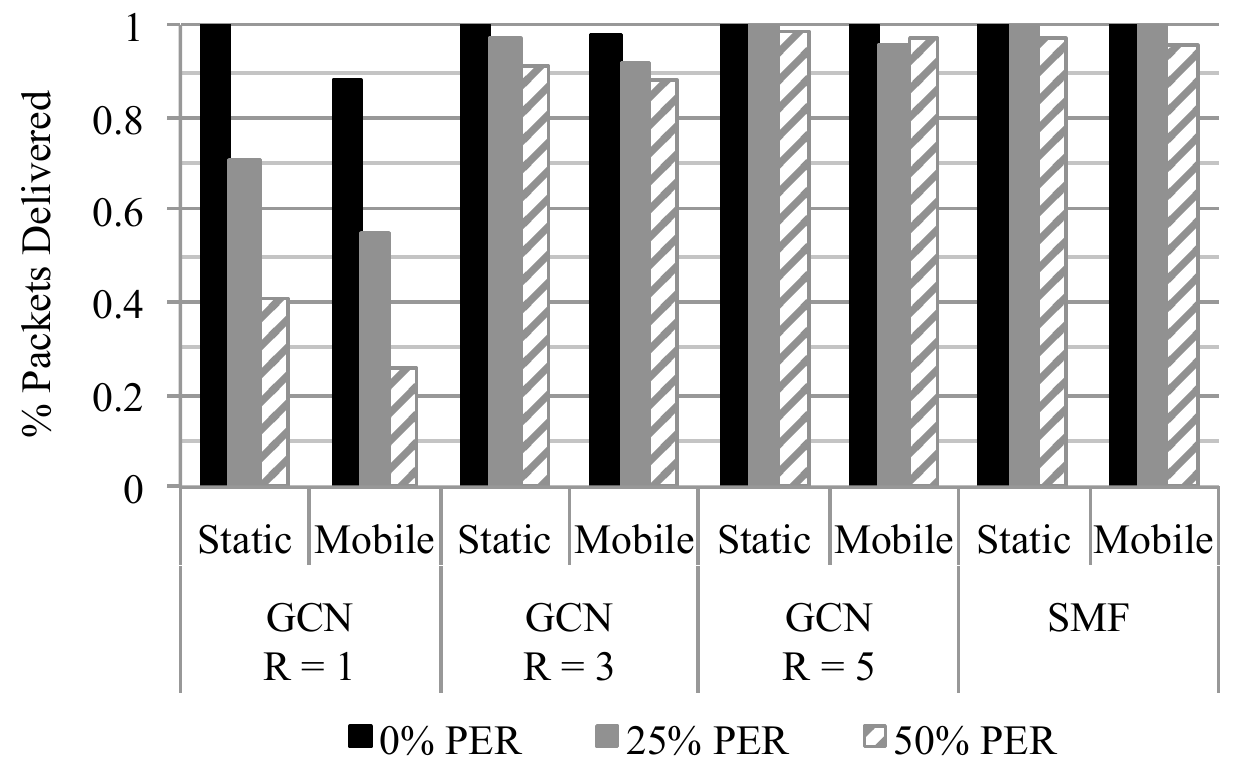} 
   \vspace{-.05in}   
   \caption{
	Packet delivery rates of GCN and SMF
   }
   \label{fig:tr_delivery}
\end{figure}
\begin{figure}[t]
   \centering
   \includegraphics[width=.4\textwidth]{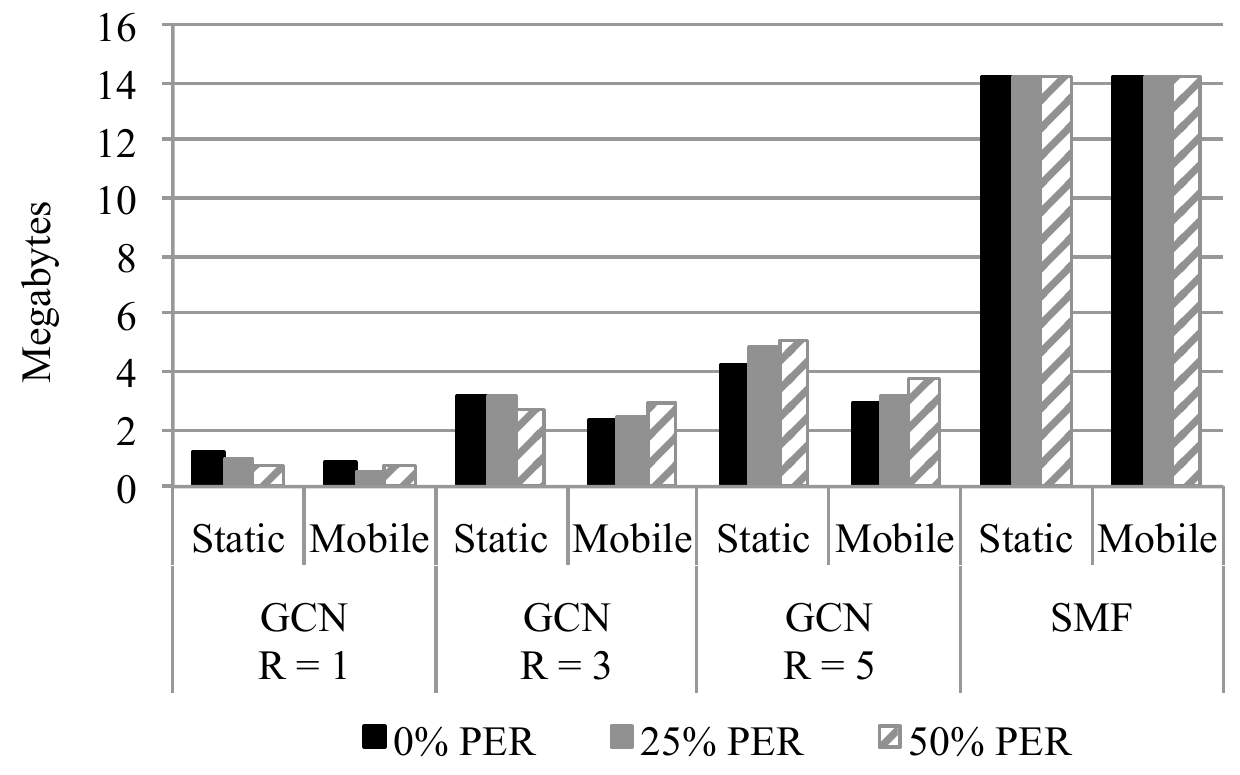} 
   \vspace{-.05in}   
   \caption{
	Bytes sent over-the-air with GCN and SMF
   }
   \label{fig:tr_sent}
  \vspace{-.2in}   
\end{figure}

\begin{figure*}[t]
    	\centering
	\subfloat{
    		\includegraphics[width=.8\textwidth]{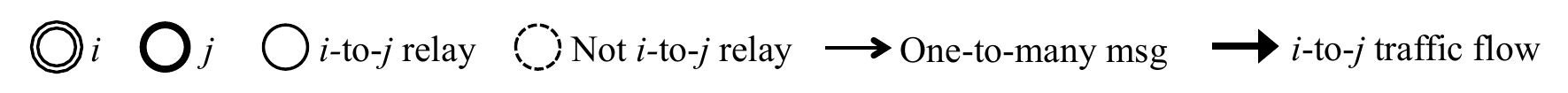} 
    		\label{fig:tf_legend}
    	} 
	\setcounter{subfigure}{0}
	\\
	\vspace{-.1in} 
    	\subfloat[User $j$ sends a message to the entire group]{
    		\includegraphics[width=0.14\textwidth]{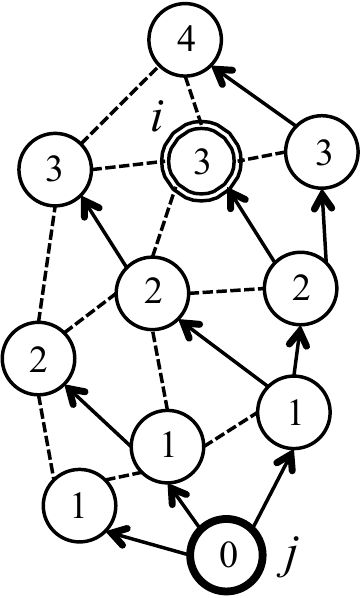} 
    		\label{fig:tf_adv}
    	}  	
	\hspace{.3in}
    	\subfloat[Only using the reverse path from $i$ to $j$]{
    		\includegraphics[width=0.14\textwidth]{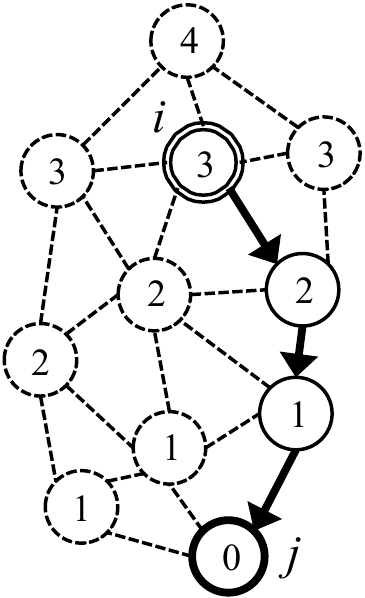} 
    		\label{fig:tf_1}
    	} 
	\hspace{.3in}	
    	\subfloat[Message path when user $i$ sets MRD to 2]{
    		\includegraphics[width=0.14\textwidth]{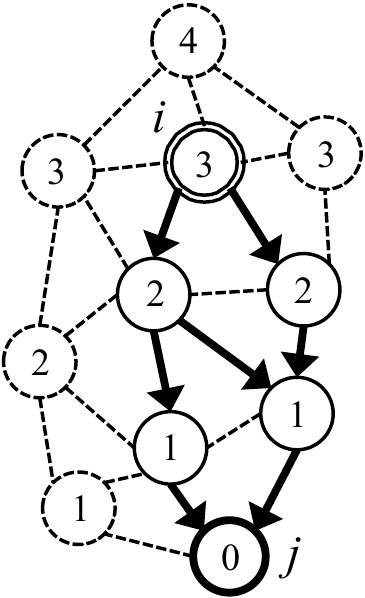} 
    		\label{fig:tf_2}
    	} 	
	\hspace{.3in}
    	\subfloat[Message path when user $i$ sets MRD to 3]{
    		\includegraphics[width=0.14\textwidth]{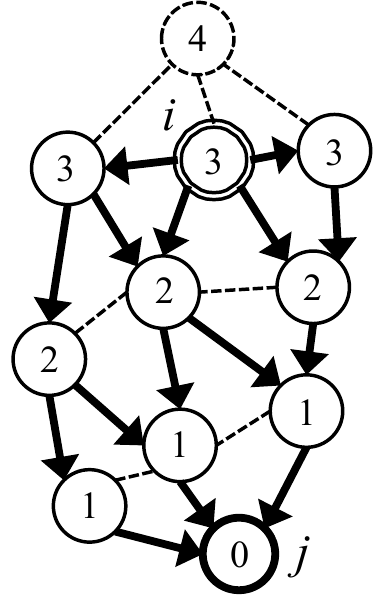} 
    		\label{fig:tf_3}
    	} 	
    	\caption{Example of a one-to-one traffic flow from user $i$ to $j$ via targeted flooding}
    	\label{fig:tf}
	\vspace{-.20in} 
\end{figure*}

The packet delivery success rate is plotted in Fig.~\ref{fig:tr_delivery}, and the total traffic sent over-the-air is plotted in Fig. \ref{fig:tr_sent}. 
For $\text{$R=1$}$, which is the case of no tunable resiliency, all packets are delivered when PER is 0\% for the static case, and 88\% of packets are delivered for the mobile case. 
With a PER of 25\% and 50\%, the number of packets delivered in the static case becomes 71\% and 41\%, respectively, and 53\% and 28\% in the mobile case, respectively. 
As the target number of data relays, $R$, increases, so does that packet delivery rate. 
Without mobility, setting $R$ to $3$ allows for a packet completion rate of 97\% for a PER of 25\%, and a completion rate of 92\% for a PER of 50\%. 
With $R=5$, all tests under all three curves have packet completion rates exceeding 98\% in the static case and 95\% in the mobile case.
With $R=5$, GCN is able to provide similar resiliency as is achieved by flooding the data across the entire region, but at a fraction of the cost.

By selectively activating additional relays throughout the group area, tunable resiliency allows for higher resiliency without consuming significant network resources. 
For SMF to achieve its high packet delivery rate, it floods the data across the entire region.
This consumes significant network resources, and does not provide substantial benefit in terms of delivery rates over GCN with tunable resiliency. 
With $R=1$, GCN uses an order of magnitude fewer network transmissions than SMF.
With a packet error rate of 50\%,  GCN with $R=5$ achieves similar packet delivery as SMF while using three times fewer over-the-air transmissions.
Additionally, we observe that for a given value of $R$, GCN maintains a fairly constant level of bytes transmitted over-the-air for the different error curves. 
This is because the selection of data relays is able to self-adjust to respond to the channel conditions, which allows it to activate a constant number of relays in presence of any packet loss that is experienced.

\vspace{-.025in} 
\subsection{Targeted Flooding}
\label{sec:ss}
\vspace{-.025in} 

In the previous two sections, we described how Group Centric Networking discovers group members and forms a resilient one-to-all communication pattern between all of them by use of tunable resiliency. 
But sending all messages to the entire group is not always efficient.
For example, a group of sensors may want to send data to a single data collector via a many-to-one traffic pattern.
Alternatively, some group member may want to query a subset of users, or have one-to-one communication with one other user.
We wish to enable these additional traffic patterns in a resilient manner without requiring additional control data to be sent throughout the network.
In this section, we present a mechanism for robustly sending data to a specific set of group members through a process we call \emph{targeted flooding}.

To be able to target transmissions towards specific users, targeted flooding uses distance information gathered from overheard packets to create a distributed gradient field towards each of the group members. 
Each transmitted packet (data or control) will be tagged with the originating user's ID, and a counter will be attached to that ID that indicates how many hops that particular packet has traveled.
Each time a packet is retransmitted, the counter is incremented. 
This adds a minimal amount of overhead to each packet. 
When a user hears a packet, it records its  distance from the originating user.
Using the distance information collected, a user can transmit a packet destined to another group member without needing to know anything about available links, or even who its own neighbors are. 
This keeps GCN's principle of not maintaining any link or neighbor information.
Each time a new message is overheard by some user, the local distance information will be updated. 
This process allows for a constant refresh of distance information without the need for dedicated control information.
Additionally, the robustness of targeted flooding is also tunable using only the distance information collected at each of the users.

In the following subsections, we describe how one-to-one, one-to-many, and many-to-many traffic patterns can be supported with GCN. 

\vspace{-.05in}

\subsubsection{One-to-One Traffic Pattern}

For some user $i$, we label its recorded distance to user $j$ as $\Delta^{i}_{j}$ (i.e., if $i$ believes it is four hops from $j$,  then $\Delta^{i}_{j}=4$).
A packet destined for a specific user will have two fields in its header: a destination, and a maximum retransmit distance (MRD). 
When a relay node hears a packet with a particular destination, it looks at that packet's MRD value, and if that value is greater than or equal to its own distance from the destination, it will rebroadcast the packet with the MRD field decremented by one.
In other words, if user $i$ receives a packet destined for $j$ with $\text{MRD} \le \Delta^{i}_{j}$, $i$ will retransmit the message with $\text{MRD} =  \Delta^{i}_{j} - 1$.

This approach will allow a packet to flood a narrow corridor towards some particular destination.
The width of the corridor that a packet travels can be modified by changing the MRD value at the originating user. 
A higher value for MRD will cause a packet to spread farther around the source, which causes a wider set of paths to be traversed as it funnels towards the destination.
Hence, the resiliency for one-to-one traffic is increased by using a higher MRD value.

We demonstrate the one-to-one flow via targeted flooding through the example in Fig. \ref{fig:tf}. 
In Fig. \ref{fig:tf_adv}, user $j$ sends a message via one-to-all to the entire group (this could be a group discovery initiated by $j$); all users learn their distance from $j$.  
In Fig. \ref{fig:tf_1}, we show the naive approach of only using the reverse path that the message traveled from $j$ to $i$; this approach relies on link state information and is vulnerable to packet loss and mobility.
In Fig. \ref{fig:tf_2}, user $i$ transmits a message destined to $j$ with the maximum retransmit distance (MRD) field set to 2; all users with a distance of two or less from $j$ retransmit the message with a MRD set to one less than their distance from $j$.
In Fig. \ref{fig:tf_3}, user $i$ transmits the message with MRD set to 3; a wider area is covered and the traffic flows across more paths from $i$ to $j$, adding additional resiliency.
Not shown is an even more resilient configuration with MRD set to 4; this will cause the top most user to participate in relaying the message, which will allow the packet to travel an even wider path. 

\vspace{-.05in}
\subsubsection{One-to-Many Traffic Pattern}

There are two forms of one-to-many traffic.
The first is where one group user desires to send a message to all of the other group members, which we typically refer to as one-to-all. 
This traffic pattern is immediately enabled after group discovery is complete, with all relay nodes retransmitting a one-to-all group  message.

The second form of one-to-many traffic is where a message is not intended for the entire group, but still has multiple destinations. 
This one-to-many traffic pattern is a straightforward extension of the one-to-one pattern. 
In particular, we wish to take advantage of any overlap between the paths a message would travel to get to different users.
For increased efficiency,  multiple destination/MRD pairs can be specified instead of having a single destination/MRD pair for a message. 
If a relay hears a message with multiple destination/MRD pairs, it follows the same process as before.
If a destination/MRD is no longer valid, then the relay simply drops that destination/MRD pair from the message before retransmitting it. 

\begin{figure}[t]
    	\centering
	\setcounter{subfigure}{0}
	\subfloat[Users learn  distance to $j$]{
    		\includegraphics[width=0.12\textwidth]{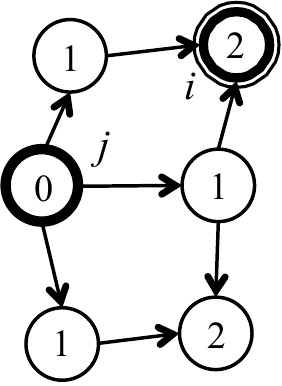} 
    		\label{fig:some_j}
    	}  	
	\hspace{.2in}
	\subfloat[Users learn  distance to $k$]{
    		\includegraphics[width=0.12\textwidth]{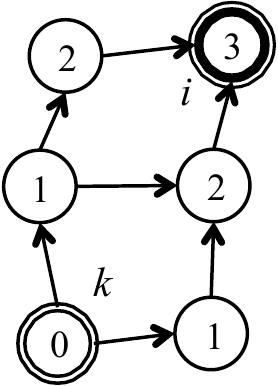} 
    		\label{fig:some_k}
    	} 
	\hspace{.2in}	
    	\subfloat[Message path from $i$ to $j$ and $k$]{
    		\includegraphics[width=0.12\textwidth]{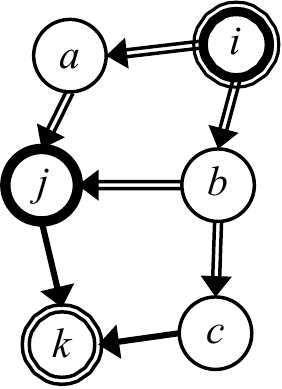} 
    		\label{fig:some_flow}
    	} 	
	\\
	\vspace{-.1in} 
	\subfloat{
    		\includegraphics[width=.45\textwidth]{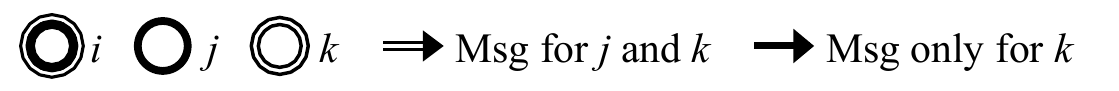} 
    		\label{fig:some_legend}
    	} 
	\vspace{-.05in} 
    	\caption{A one-to-many traffic flow from user $i$ to both $j$ and $k$}
    	\label{fig:some}
	\vspace{-.2in} 
\end{figure}

An example of one-to-some traffic pattern is shown in Fig. \ref{fig:some}. 
User $i$ has a packet that it wishes to send to both $j$ and $k$. 
In Fig. \ref{fig:some_j}, $j$ sends a one-to-many message, and all users in the group learn their distance to $j$.
Similarly in Fig. \ref{fig:some_k}, all users learn their distance to $k$ from a one-to-many message that $k$ sent.
In Fig. \ref{fig:some_flow}, user $i$ sends a message destined to both $j$ and $k$. 
The packet has two destination/MRD fields: 
the first is set to $j$/1, and the second is set to $k$/2.
Relays $a$ and $b$ retransmit the message and set the MRD field to 0 and 1 for $j$ and $k$, respectively. 
Relay $c$ receives the message and sees that the MRD field for destination $j$ is no longer valid; $c$ drops $j$ and resends the packet with only $k$ as a destination and an MRD value set to 0. 
User $j$ will also drop itself as a destination from the message before retransmitting.

\subsubsection{Many-to-Many Traffic Pattern}

The many-to-many pattern is implemented as a collection of one-to-many traffic patterns operating jointly. 
Numerous efficiencies can be gained in the many-to-many traffic pattern by performing coordinated data fusion, source coding, or network coding between the various users and traffic flows~\cite{pradhan2003distributed,xiong2004distributed,pattem2008impact,katti2006xors}.
Applying these techniques within GCN is a topic of future study.

\subsubsection{Targeted Flooding Performance Evaluation}

We evaluate the performance of targeted flooding by running the following simulation in NS3.
100 users are uniformly distributed in a circular region with a radius of 100 meters. 
Any user in this region can be a group member with a probability of 25\%. 
We test the following three packet error rates (PER): 0\%, 25\%, and 50\%. 
Users are either stationary or move according the  random waypoint model where each user selects a speed from a uniform distribution between 0 and 5 m/s.
One group member is selected at random to initiate group discovery; this user is labeled the source. 
Each group member sends a one-to-one message to the source once per second for 100 seconds.
The source sends an empty data packet to the group once every two seconds to allow for updated distance information. 
The number of desired data relays $R$ is set to 5.

Three different resiliency values are used for the one-to-one flow: low, medium, and high.
For low resiliency, the MRD is set to one less than that user's distance to its intended destination; i.e., if user $i$ is sending a packet to user $j$, and user $i$ has distance value $\Delta^{i}_{j} = d$, then with low resiliency, the MRD would be set to $d-1$.
Medium resiliency has the MRD set to be the same as that user's distance to the destination, and with high resiliency, the MRD is set to one greater than that user's distance.
In addition to comparing against SMF, which floods the data across the entire region, we compare against AODV, which is potentially more comparable since it finds a one-to-one unicast path between users. 
We measure the number of packets successfully received at each group member and the total number of bytes sent over-the-air.
Fifty random tests are run. 

\begin{figure}[t]
   \centering
   \includegraphics[width=0.40\textwidth]{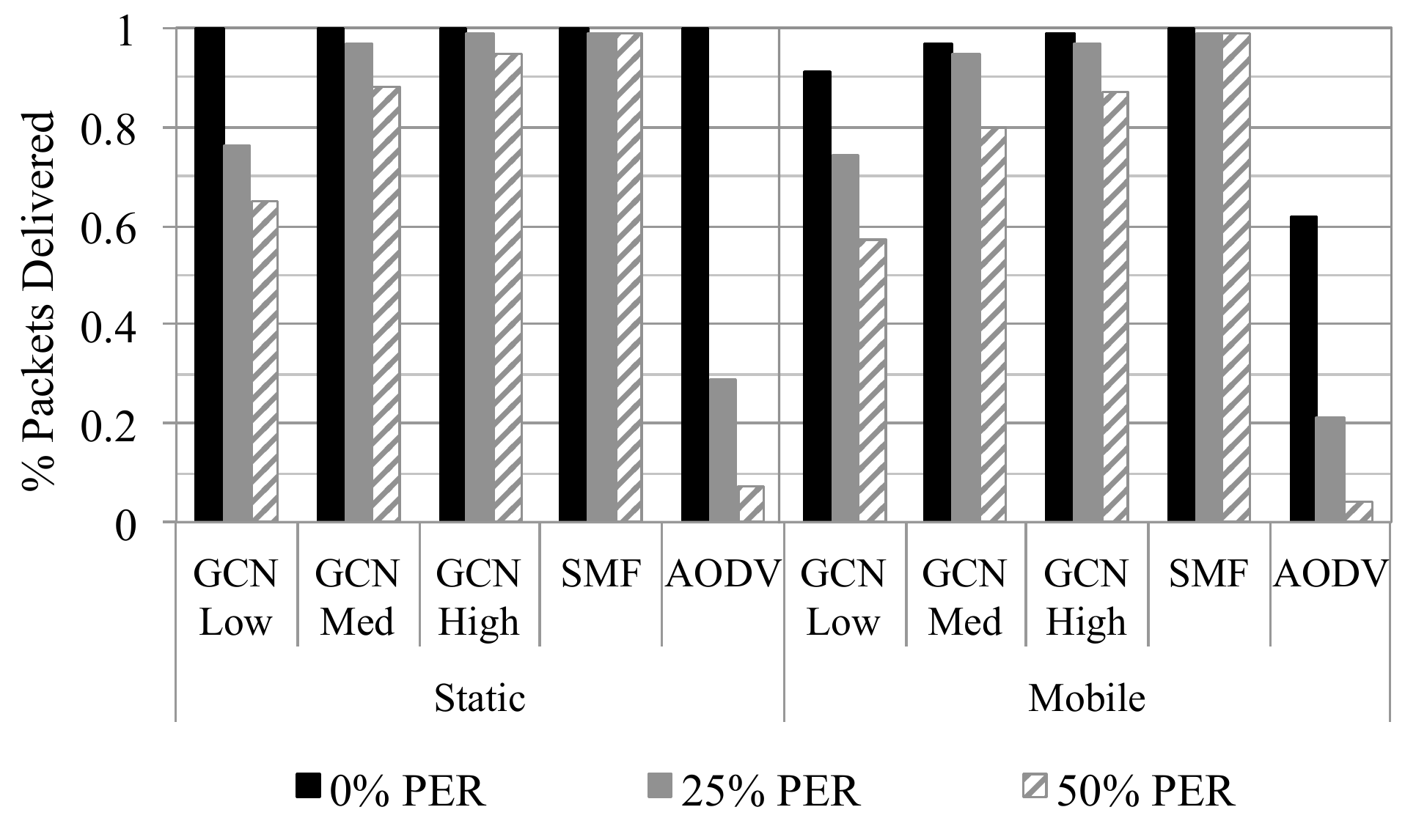} 
   \vspace{-.1in}   
   \caption{Packet delivery rates using targeted flooding}
   \label{fig:tf_delivery}
   \vspace{-.1in}
\end{figure}

\begin{figure}[t]
   \centering
   \includegraphics[width=.40\textwidth]{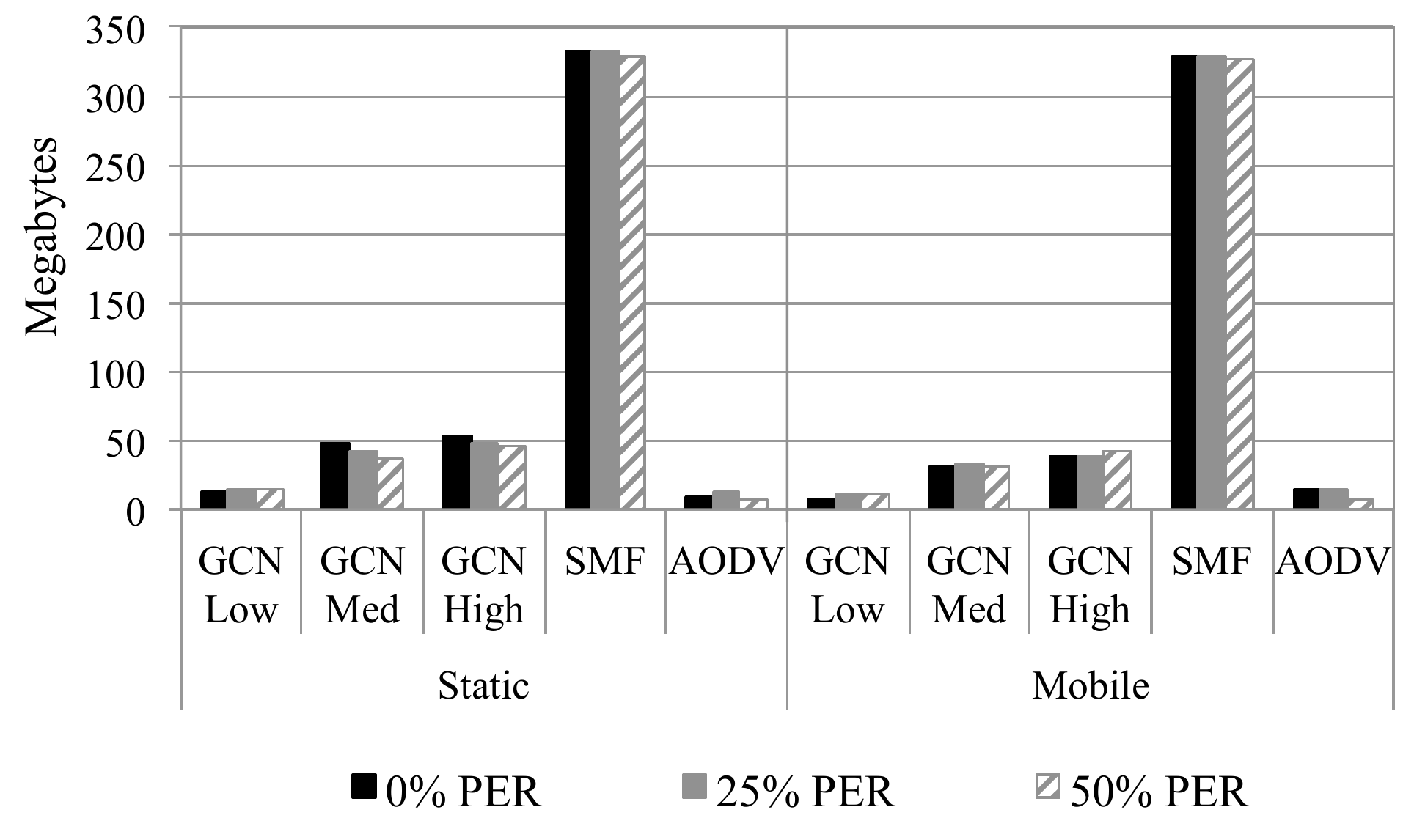} 
   \vspace{-.1in}   
   \caption{Bytes  over-the-air  using targeted flooding}
   \label{fig:tf_sent}
  \vspace{-.1in}   
\end{figure}

Fig. \ref{fig:tf_delivery} shows the percent of packets successfully received at the source, and Fig. \ref{fig:tf_sent} shows the number bytes transmitted over-the-air to do so for each of the schemes tested.
SMF is able to deliver close to 100\% of the packets, but at great cost with respect to network resource utilization. 
Conversely, AODV transmits much less over-the-air, but is unable to successfully deliver packets in a lossy environment. 
In the static case, AODV delivers 29\% of packets when the PER is 25\%, and has only a 7\% delivery when the PER is 50\%. 
In the mobile case, AODV only delivers 62\% of packets with a PER of 0\%, and only delivers 4\% of packets when the PER is 50\%. 
GCN is able to offer the resiliency of flooding at a significantly lower network load. 
For the static case with medium resiliency, the packet delivery rate is 99\% for a PER of 25\%. 
With high resiliency for the static case, packet delivery is 95\% for a PER of 50\%.
With mobility, high resiliency has a delivery rate of 99\%, 96\%, and 87\% for a PER of 0\%, 25\%, and 50\%, respectively. 
GCN's bytes transmitted over-the-air with low resiliency is approximately an order of magnitude lower than SMF, and actually is lower than AODV in the mobile case. 
GCN with medium and high resiliency uses between six to eight times fewer resources than SMF.

\section{Implementation and Evaluation}
\label{sec:gcn_implementation}

In this section, we overview our implementation of Group Centric Networking and perform an evaluation using our implementation of all of the components of GCN working together. 

\vspace{-.05in}
\subsection{Implementation}

We implemented the full GCN protocol in both simulation and emulation. 
The reason for implementing across multiple environments is to allow us to verify results across multiple platforms. 
This enables us to have confidence that GCN performs as expected.

To operate in simulation (NS3), we leverage NS3 Direct Code Execution (DCE)~\cite{ns3dce}, which provides a framework to execute existing implementations of userspace and kernelspace network protocols with minimum source code changes. 
For emulation testing, we leveraged the Extendable Mobile Ad-hoc Network Emulator (EMANE) \cite{emane} that emulates layers 1 and 2 (radio and link layers) of the network stack in real-time, and the Common Open Research Emulator (CORE)~\cite{core_emane} to help configure, launch, and execute real-time experiments. 
CORE creates Linux containers that represent independent network users and configures network interfaces, access lists, and processes (which includes the GCN layer). 
To date, we've successfully validated GCN operation on a 300 node emulation network emulating dozens of hardware platforms.


\vspace{-.025in}
\subsection{Evaluation of GCN}
\vspace{-.025in}

In this section, we examine the resiliency and scalability of GCN, and compare GCN against SMF and AODV.
In particular, we measure the packet delivery rate and the total number of bytes sent over-the-air. 
All tests are performed via simulation using NS3 DCE. 

For our tests, we vary the following parameters: number of users, number of group members, mobility, and packet error rate.
Users are uniformly distributed in a circular region with a radius of 100 meters.
We test both 50 and 100 users in the network, 
where a user can be a group member with a probability $P_g$ of either 10\% or 25\%. 
We test both a static and mobile network. 
For mobility, users move according the  random way point model with zero hold time and a speed that is uniformly selected  between 0 to 5 m/s.

\begin{figure}[t]
	\centering
	\includegraphics[width=0.3\textwidth]{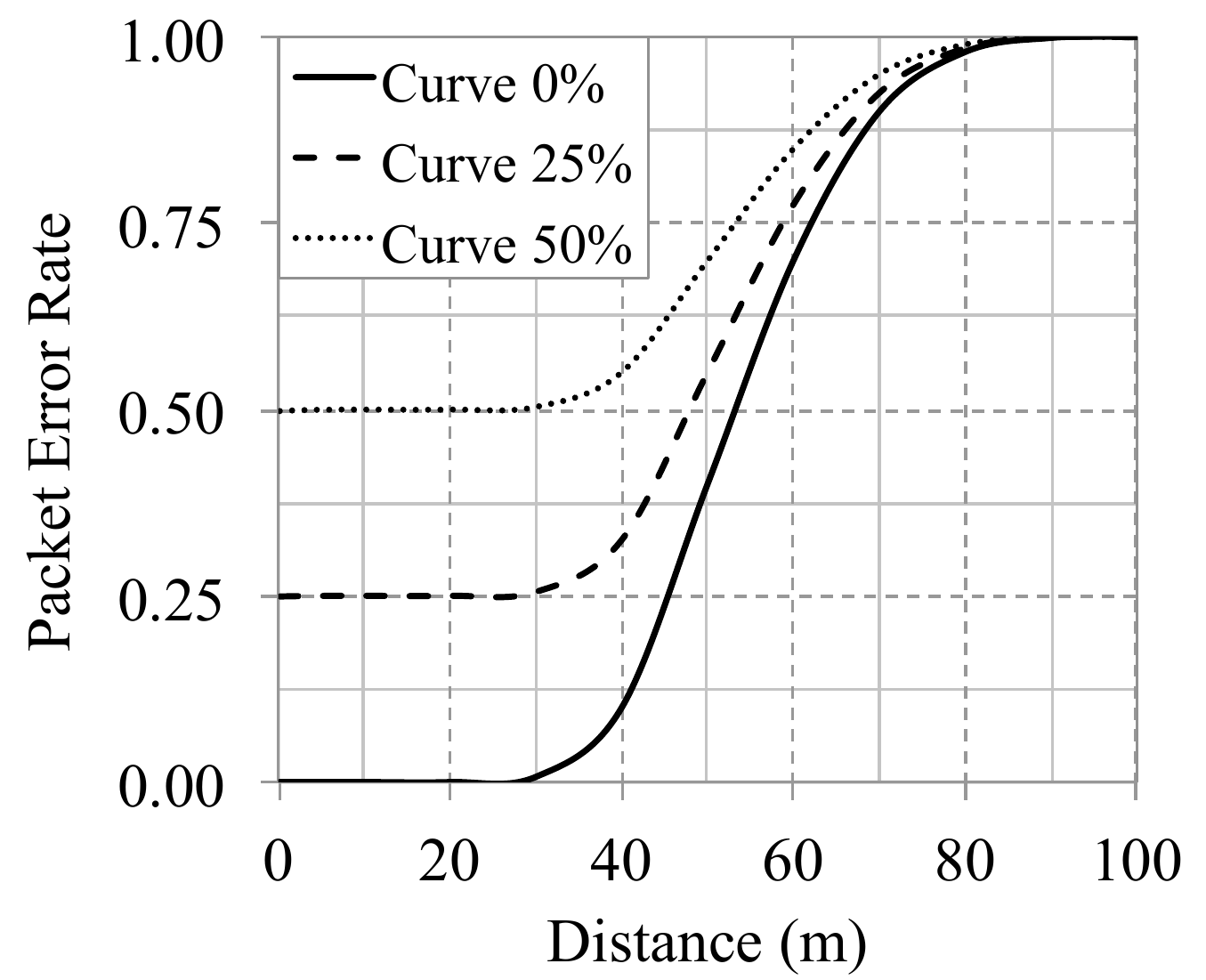}
	\vspace{-.1in}
	\caption{Packet error rate for IEEE 802.15.4 devices \cite{petrova2006performance}}
	\label{fig:PERcurve}
	\vspace{-.25in}
\end{figure}

In order to better evaluate a realistic environment where these smart-devices will be deployed, 
 for our wireless channel model we use the packet error rate (PER) curve for IEEE 802.15.4 devices from \cite{petrova2006performance}, which is reproduced as \texttt{Curve 0\%} in Fig. \ref{fig:PERcurve}. 
The packet error rates in \cite{petrova2006performance} were determined through both simulations and hardware measurements.
This PER curve assumes no loss for short range transmissions; however, in the presence of interfering wireless devices, one would not expect 0\% packet loss for transmissions at close range.
Various papers have tried to quantify the effects of interference on packet reception rates for devices operating in the 2.4 GHz ISM band (where 802.15.4 operates) \cite{petrova2006performance,Bandspeed,hithnawi2014understanding}.
These studies find that loss can be on the order of 25\% or greater.
We define two new curves for a higher loss environment where the minimum PER is either 25\% or 50\% for short range transmissions.
The PER curve to model interference that causes 25\% packet loss is constructed by multiplying the packet success rate at any given distance $d$ (i.e., $(1-\text{PER}(d))$) for \texttt{Curve~0\%}  of Fig. \ref{fig:PERcurve} by  $(1-0.25)$.  
The PER curve for 50\% packet loss is constructed in a similar fashion.
These two curves are also shown in Fig. \ref{fig:PERcurve}, and are labeled  \texttt{Curve~25\%}, and  \texttt{Curve~50\%}.

The traffic for all scenarios is as follows. 
A group member is randomly selected as the source, and the source node initiates the group discovery process.
The source node then transmits one message per second to all other group members via a one-to-many data pattern. 
All other group members transmit a packet via a one-to-one transmission back to the source node once per second for 100 seconds. 
The same traffic pattern is run using GCN, SMF, and AODV. 
Similar to our previous tests, the minimum TTL is selected for SMF such that every group member can reach every other group member, and AODV is run with its default parameters. 

We consider three different resiliency levels for GCN: low, medium, and high, where the number of desired data relays $R$ is set to either 3, 6, or 9, respectively.
Recall that the parameter $R$ sets the number of additional relays that are selected during the group discovery process. 
For all three resiliency levels, the maximum retransmit distance (used for the many-to-one traffic pattern) is set to be one greater than a user's distance to its intended destination.


\begin{figure}[t]
    	\centering
    	\subfloat[50 users]{
    		\includegraphics[width=0.4\textwidth]{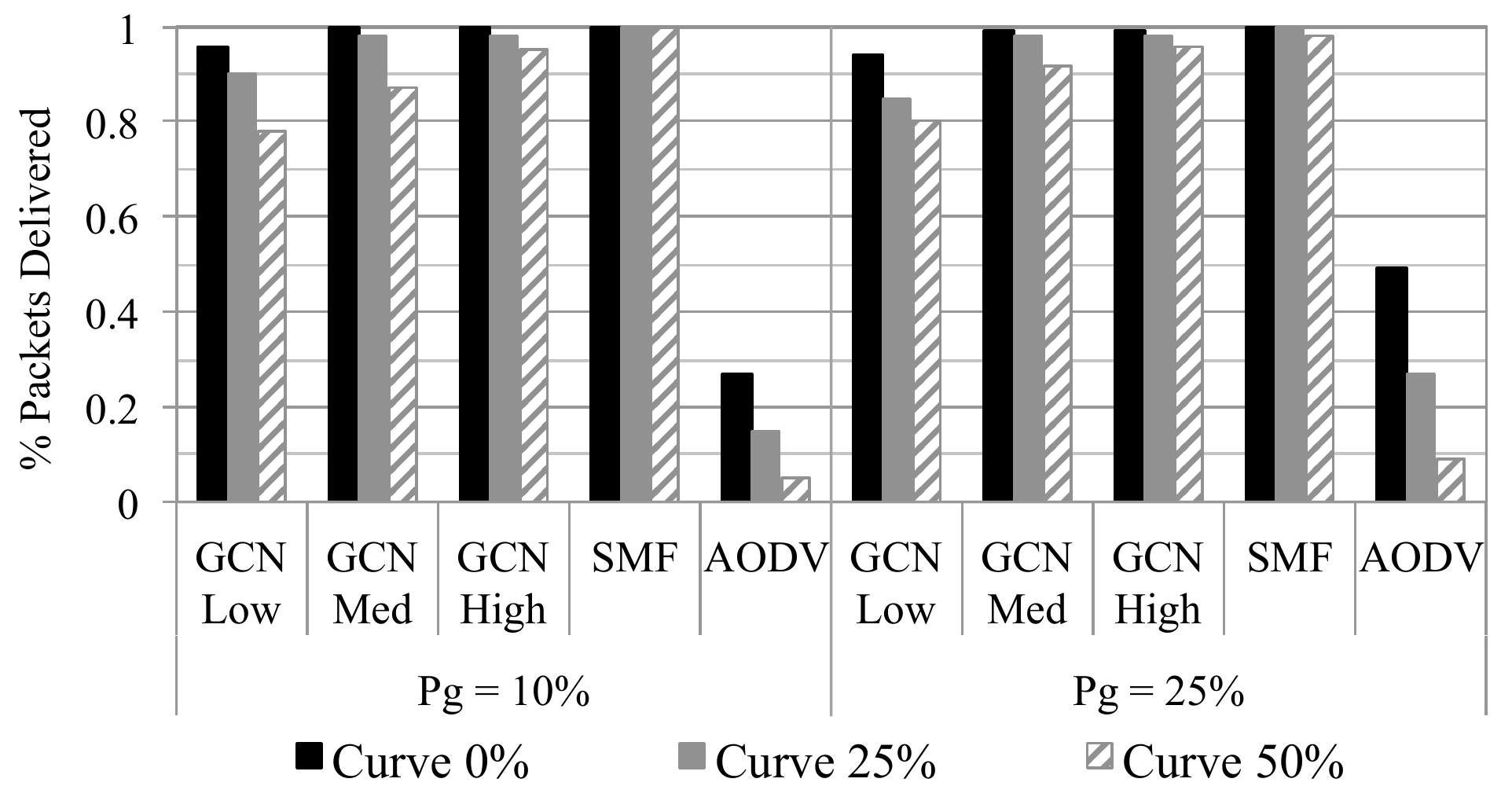} 
    		\label{fig:test_delivery_static_50}
    	}  	
	\\
	\vspace{-.1in}
    	\subfloat[100 users]{
    		\includegraphics[width=0.4\textwidth]{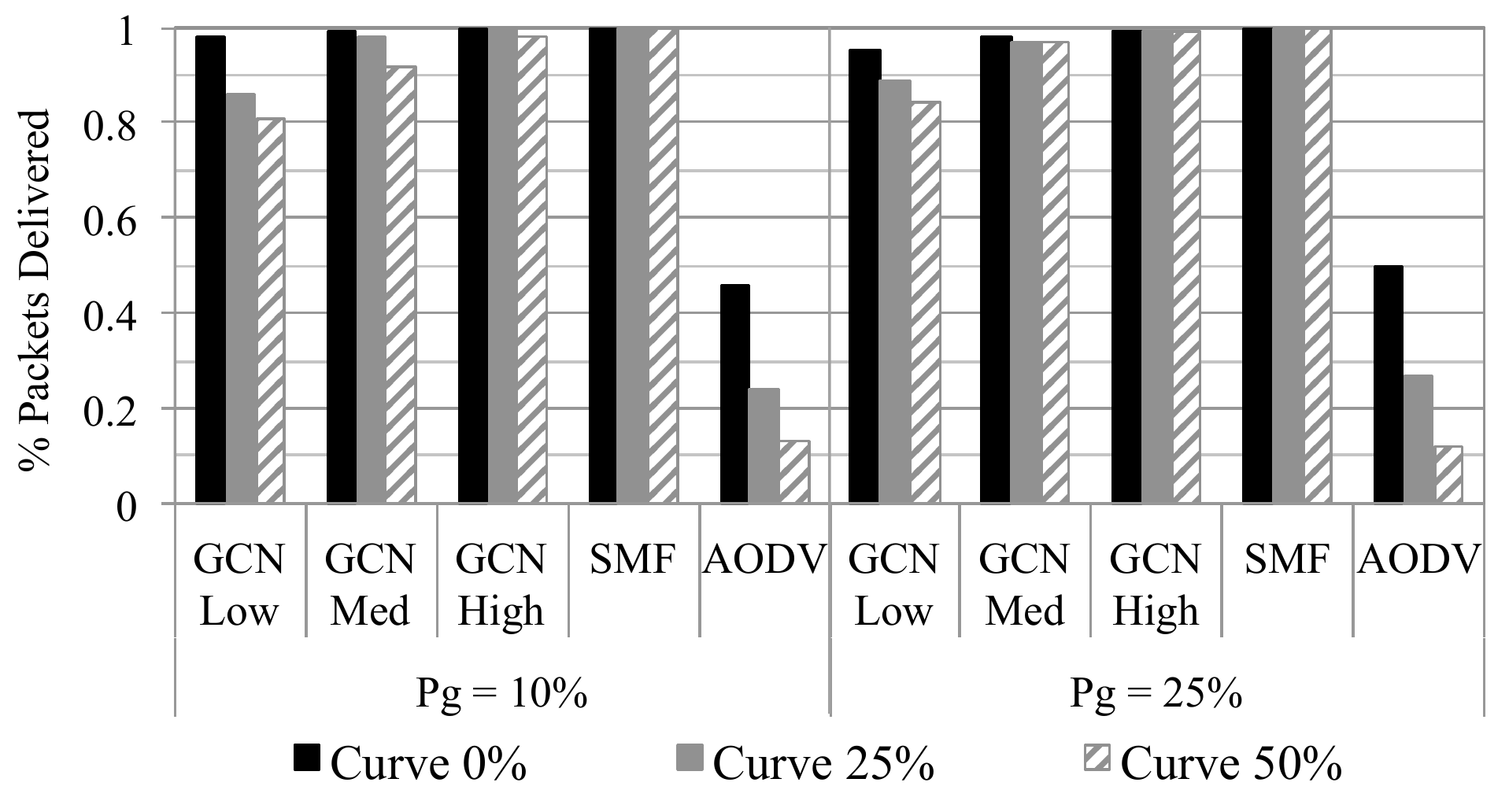} 
    		\label{fig:test_delivery_static_100}
    	} 
	\vspace{-.1in} 
    	\caption{Static network: packet delivery rate}
    	\label{fig:test_delivery_static}
	\vspace{-.2in} 
\end{figure}

\begin{figure}[t]
    	\centering
    	\subfloat[50 users]{
    		\includegraphics[width=.4\textwidth]{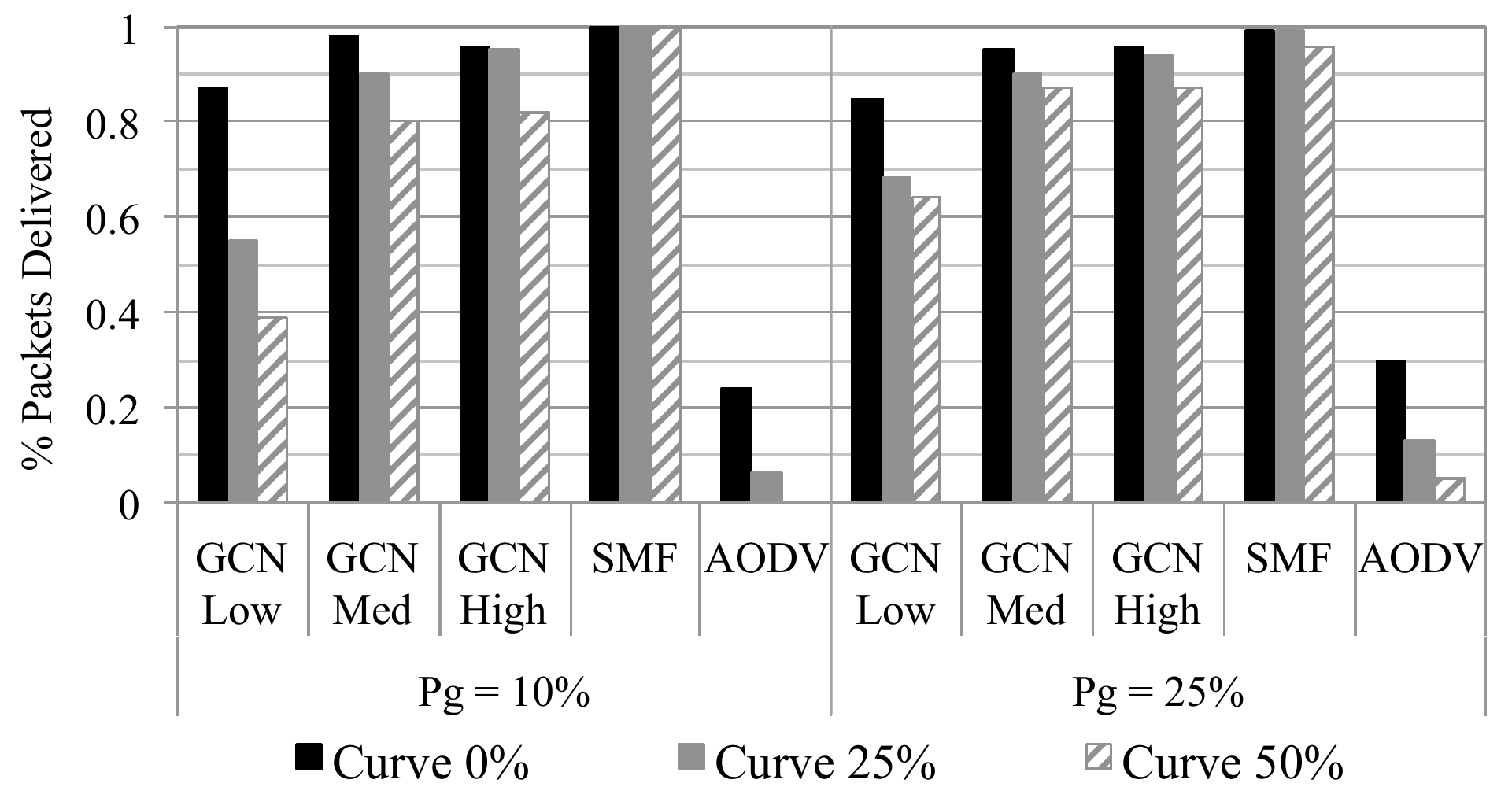} 
    		\label{fig:test_delivery_mobile_50}
    	}  	
	\\
	\vspace{-.1in}
    	\subfloat[100 users]{
    		\includegraphics[width=.4\textwidth]{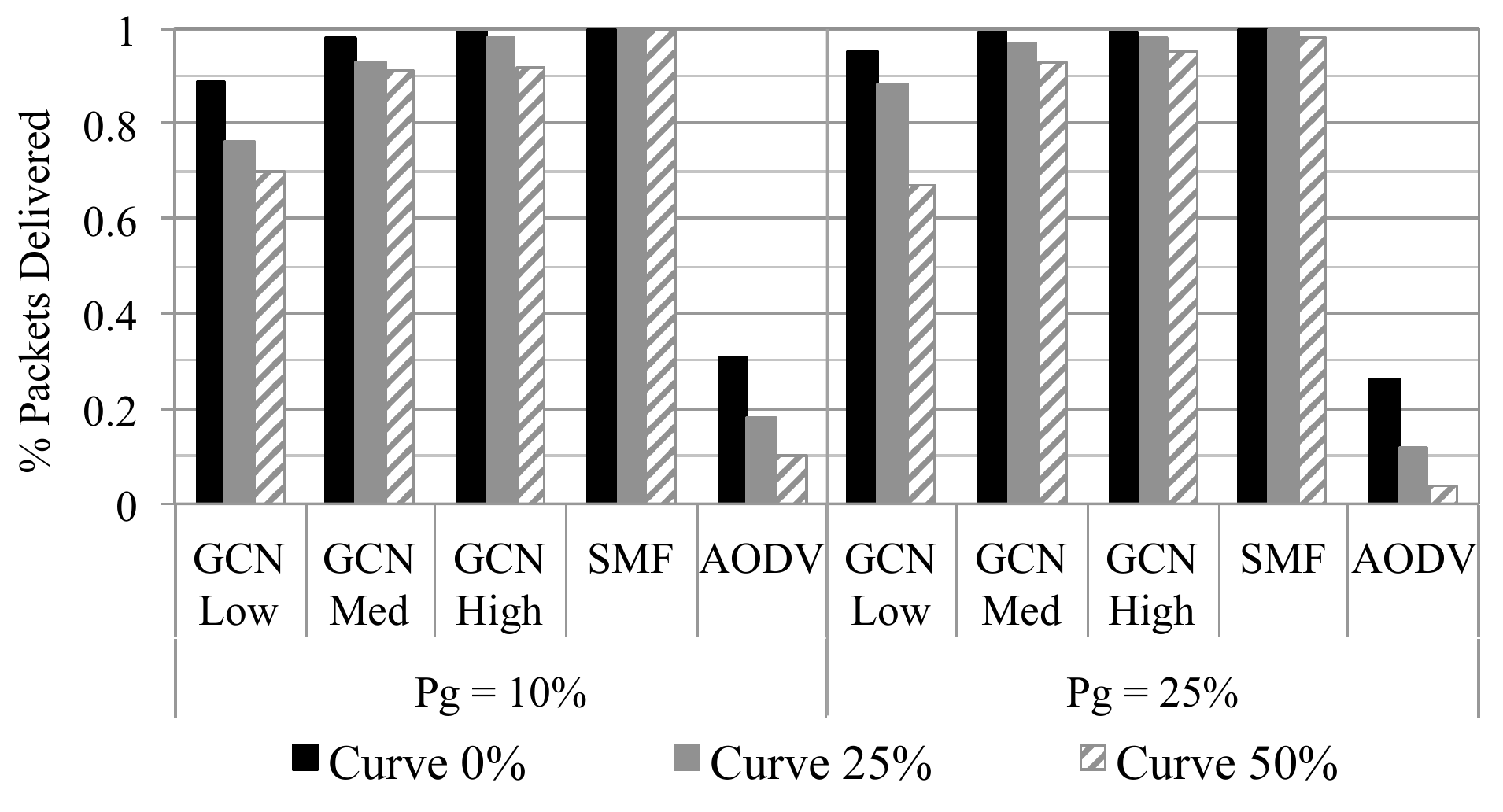} 
    		\label{fig:test_delivery_mobile_100}
    	} 
	\vspace{-.1in} 
    	\caption{Mobile network: packet delivery rate}
    	\label{fig:test_delivery_mobile}
	\vspace{-.2in} 
\end{figure} 

Fig. \ref{fig:test_delivery_static} shows the packet delivery rate for a static network. 
For GCN with low resiliency, approximately 95\% of packets are delivered under \texttt{Curve 0\%} for all combinations of network and group size. 
This approaches the delivery success rate of SMF, which floods a packet across the network.
For \texttt{Curve~25\%}, GCN is able to deliver 97\% of packets using medium resiliency, 
and under \texttt{Curve~50\%}, which has a baseline packet error rate of 50\%, GCN with high resiliency is able to deliver over 95\% of packets for all cases tested.
In contrast, AODV is only able to successfully deliver 28\% to 50\% of packets under \texttt{Curve~0\%}, and only 6\% to 12\% under  \texttt{Curve~50\%}.
The reason for AODV's poor performance under the relatively benign \texttt{Curve~0\%} is as follows.
In \texttt{Curve~0\%}, short links are error-free and longer links have high error rate. 
AODV builds a shortest path route  by using the set of exchanged hello messages between users of the network. 
With sufficiently high frequency, hello messages are successfully exchanged across high error links, and since these links are of longer distance, these poor quality links get used to build shortest path routes. 

\begin{figure}[t]
    	\centering
    	\subfloat[50 users]{
    		\includegraphics[width=0.39\textwidth]{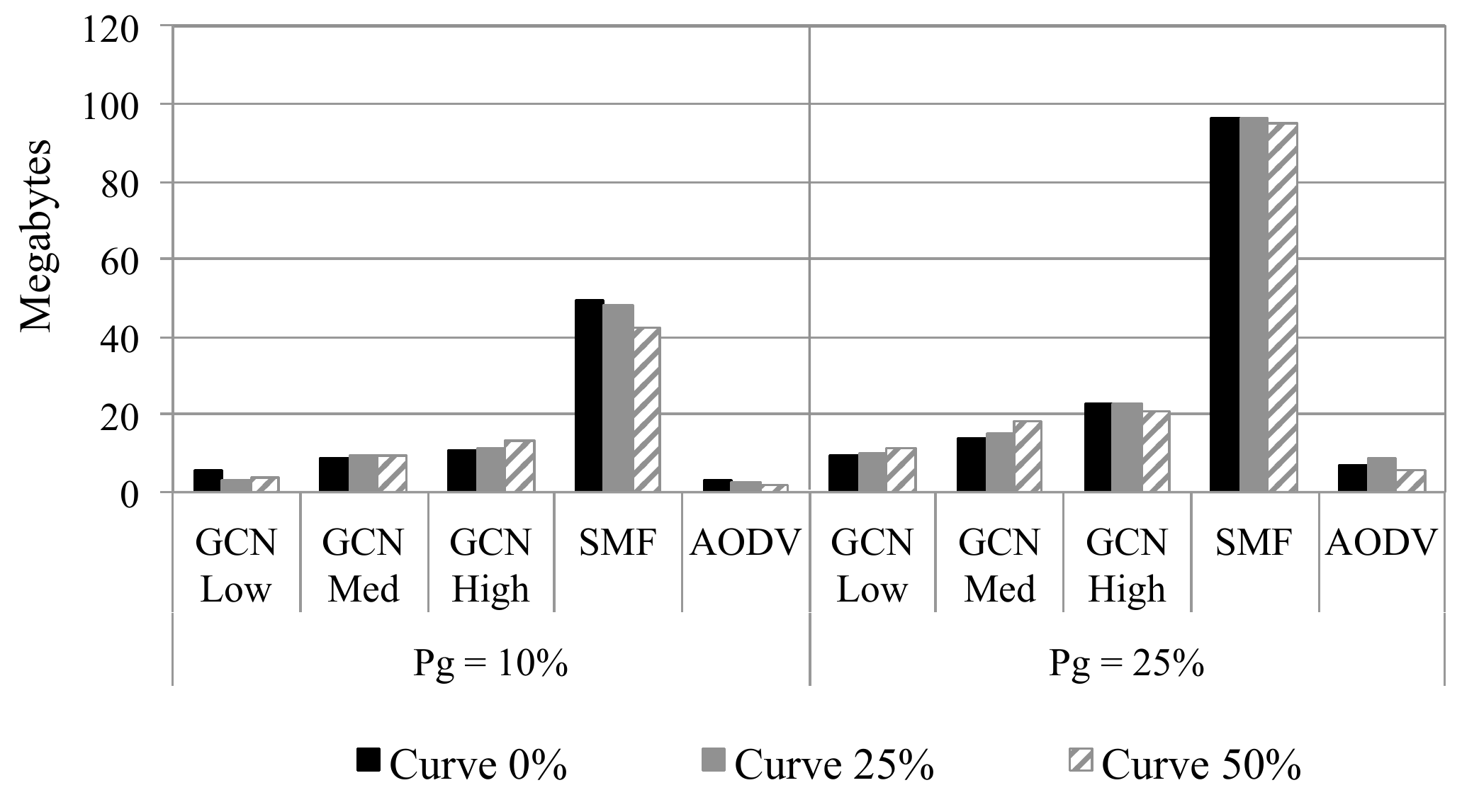} 
    		\label{fig:test_tx_50_static}
    	}  	
	\\
	\vspace{-.15in} 
    	\subfloat[100 users]{
    		\includegraphics[width=.39\textwidth]{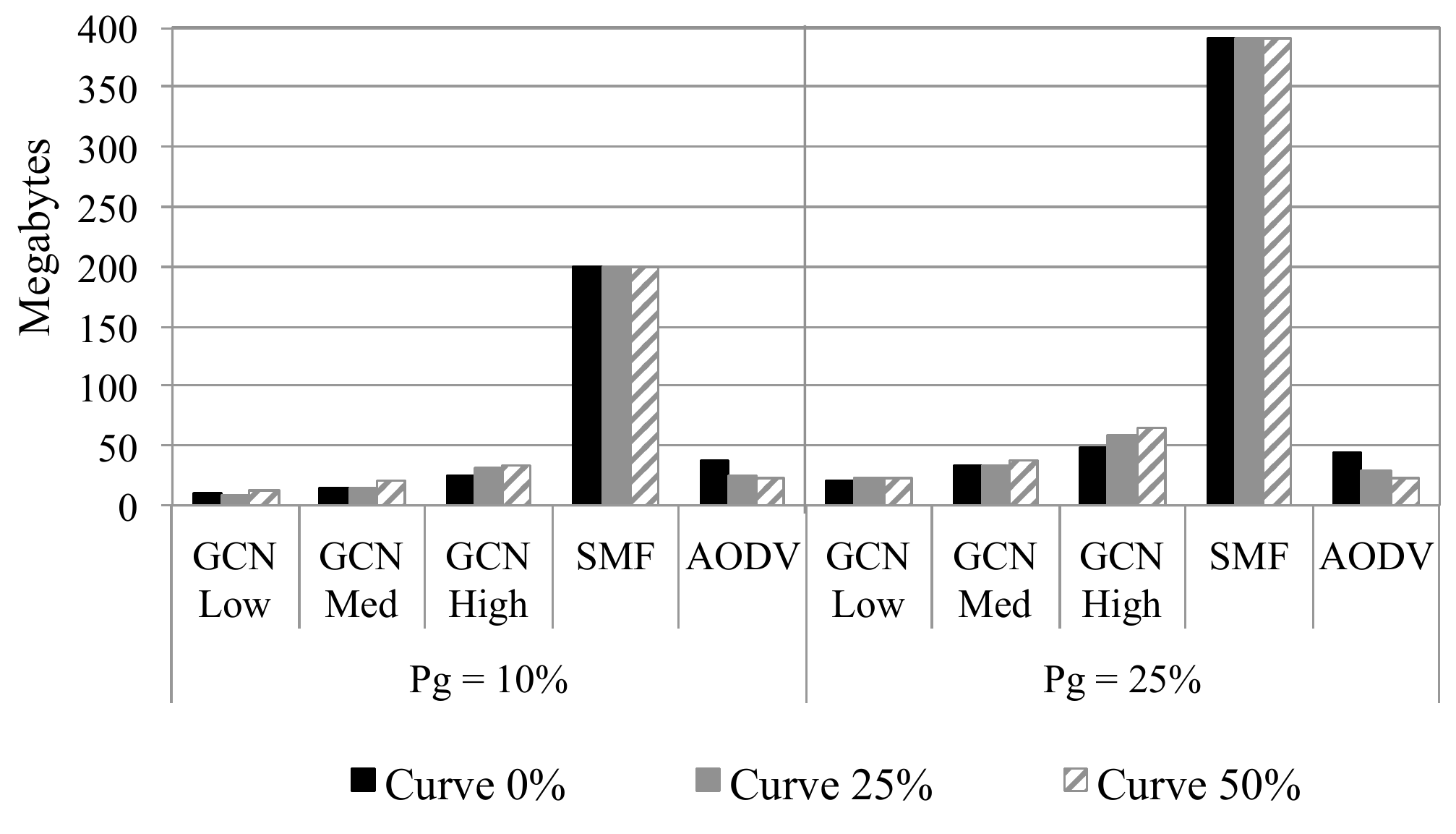} 
    		\label{fig:test_tx_100_static}
    	} 
	\vspace{-.1in} 
    	\caption{Static network: over-the-air transmissions}
    	\label{fig:test_tx_static}
	\vspace{-.2in} 
\end{figure}


Fig. \ref{fig:test_delivery_mobile} shows the packet delivery rate for networks with mobile users. 
As expected, packet delivery rates are lower for all cases tested. 
AODV now reaches only a maximum of 31\% delivery, and 
goes as low as 0\% delivery for the case of 50 users, $P_{g}=10\%$, and \texttt{Curve~50\%}. 
GCN with high resiliency in the 50 user network is able to deliver over 82\% of packets  under \texttt{Curve~50\%}, and delivers over 95\% under  \texttt{Curve~0\%}.
In the 100 user network, GCN with high resiliency delivers over 91\% of packets under \texttt{Curve~50\%}, and delivers almost 100\% under  \texttt{Curve~0\%}.

We note that GCN under low resiliency for \texttt{Curve~25\%} and \texttt{Curve~50\%} has has poorer performance for smaller networks and lower group sizes (i.e., 50 users and $P_{g}=10\%$).
This is because when there are few users, coverage of the local area under low resiliency is insufficient to adequately provide connectivity for all users in the presence of mobility.
Using medium resiliency significantly improves performance for smaller networks and lower group sizes under \texttt{Curve~25\%} and \texttt{Curve~50\%}.


While packet delivery is the purpose for any network protocol, future networks of power and bandwidth constrained smart-devices must be able to reliably deliver packets using as few transmissions as possible. 
Fig. \ref{fig:test_tx_static} 
shows the bytes transmitted over-the-air to deliver the traffic sent  for the static scenario. 
The mobile results are excluded due to space constraints, but are similar to the results from the static case.
While SMF was the most reliable of the  different approaches tested, it came at a very high cost. 
SMF floods each packet across the network, and hence transmits a very large number of messages throughout the network. 
This is particularly inefficient for the one-to-one and many-to-one traffic pattern, where data will be rebroadcast in areas of the network far from the destinations. 
Additionally, areas with a large number of users will have the same message retransmitted more times than was necessary to have all users receive the packet.

In contrast, GCN is able to achieve delivery rates comparable to flooding while using an order of magnitude fewer network resources. 
GCN is able to selectively choose how many users will relay data in any given area, and is able to keep that number of users relatively constant regardless of the packet error rate being experienced.
Furthermore, the one-to-one traffic pattern uses targeted flooding to reliably transmit a packet towards its intended destination, as opposed to SMF that causes each packet to be flooded throughout the entire network.
GCN allows for highly resilient communication without utilizing significant network resources.
AODV and GCN utilize a comparable amount of network resources, but as was shown earlier, AODV is unable to reliably deliver packets in a lossy network. 
Under \texttt{Curve 50\%}, AODV had delivery rates ranging from 0\% to 12\%, while GCN had delivery rates ranging from 82\% to 100\%. 
GCN is able to successfully deliver data in a lossy mobile environment similar to flooding without incurring the high cost of flooding. 

\section{Conclusion}
\label{sec:conclusion}

In this paper, we introduce Group Centric Networking (GCN), which is designed to provide resilient and scalable multi-hop wireless communications for emerging networks of smart-devices.
We anticipate that networks of these devices will operate collaboratively as a group in a local region.
Since many of these devices will be resource limited and will be operating in a lossy environment,
GCN is designed to enable these devices to operate collaboratively in a highly efficient and resilient fashion, while not sacrificing the users' ability to communicate with one another. 

We do a full protocol implementation of GCN in NS3, and verify GCN in emulation. 
We find that GCN utilizes up to an order of magnitude fewer network resources than traditional wireless networking schemes, while also achieving high connectivity and resiliency. 

We are currently continuing with development for GCN, with areas of study including: 
using GCN in a multi-channel system, using GCN with systems of directional smart-antennas, additional approaches for resiliency, porting GCN to hardware devices, and using GCN to support various network applications. 

\bibliographystyle{IEEEtran}
\footnotesize
\bibliography{bib}

\end{document}